\documentclass[prc,amsmath,twocolumn,showkeys,showpacs,superscriptaddress]{revtex4}

\usepackage{gensymb}
\usepackage{graphicx,color}
\usepackage{amssymb}
\usepackage{enumerate}
\usepackage{verbatim}
\usepackage{natbib}
\usepackage{amsmath}

\begin{document}
  \newcommand {\nc} {\newcommand}
  \nc {\Sec} [1] {Sec.~\ref{#1}}
  \nc {\IR} [1] {\textcolor{red}{#1}} 
  \nc {\IB} [1] {\textcolor{blue}{#1}}
  \nc {\IG} [1] {\textcolor{green}{#1}}

\title{Uncertainty Quantification for Optical Model Parameters}

\author{A.~E.~Lovell}
\affiliation{National Superconducting Cyclotron Laboratory, Michigan State University, East Lansing, MI 48824, USA}
\affiliation{Department of Physics and Astronomy, Michigan State University, East Lansing, MI 48824, USA}
\author{F.~M.~Nunes}
\affiliation{National Superconducting Cyclotron Laboratory, Michigan State University, East Lansing, MI 48824, USA}
\affiliation{Department of Physics and Astronomy, Michigan State University, East Lansing, MI 48824, USA}
\author{J.~Sarich}
\affiliation{Mathematics and Computer Science Division, Argonne National Laboratory, Lemont, IL 60439, USA}
\author{S.~M.~Wild}
\affiliation{Mathematics and Computer Science Division, Argonne National Laboratory, Lemont, IL 60439, USA}

\date{\today}

%%%%%%%%%%%%%%%%%%%%%%%%%%%%%%%%%%%%%%%%%%%%%%%%%%%%%%%%%%%%%%%%%%%%%%%%%%%%%%%%%%%%%%%%%%%%%%%%%%%%%%%%%%%%%%%%%%%%%%%%%%%%%%%%%%%

\begin{abstract}
\begin{description}
\item[Background:]  Although uncertainty quantification has been making its way into nuclear theory, these methods have yet to be explored in the context of reaction theory.  For example, it is well known that different parameterizations of the optical potential can result in different cross sections, but these differences have not been systematically studied and quantified.
\item[Purpose:]  The purpose of this work is to investigate the uncertainties in nuclear reactions that result from fitting a given model to elastic-scattering data, as well as to study how these uncertainties propagate to the inelastic and transfer channels.
\item[Method:]  We use statistical methods to  determine a best fit and create corresponding 95\% confidence bands. A simple model of the process is fit to elastic-scattering data and used to predict either inelastic or transfer cross sections.  In this initial work, we assume that our model is correct, and the only uncertainties come from the variation of the fit parameters.
\item[Results:]  We study a number of reactions involving neutron and deuteron projectiles with energies in the range of 5--25 MeV/u, on targets with mass $A$=12--208. 
We investigate the correlations between the parameters in the fit.  The case of deuterons on $^{12}$C is discussed in detail:  the elastic-scattering fit and the prediction of $^{12}$C(d,p)$^{13}$C transfer angular distributions, using both uncorrelated and correlated $\chi^2$ minimization functions. The general features for all cases are compiled in a systematic manner to identify trends. 
\item[Conclusions:]  Our work shows that, in many cases, the correlated $\chi ^2$ functions (in comparison to the uncorrelated $\chi^2$ functions) provide a more natural parameterization of the process.  These correlated functions do, however, produce broader confidence bands. Further optimization may require improvement in the models themselves and/or more information included in the fit.
\end{description}
\end{abstract}



\maketitle

\section{Introduction}

%\begin{itemize}
%\item Why is reaction theory important
%\item Why is UQ needed in reaction theory
%\item Previous methods of uncertainty quantification - comparisons between models/parameterizations
%\item Current applications in the field - Bayesian in EFT and elsewhere
%\item Current applications outside of the field - HEP, solar masses, chemistry, geology, etc.
%\item Motivation - differences between models/parameterizations, differences in transfer cross sections from the same elastic, fitting with lack of angular coverage, etc.
%\end{itemize}

For nuclei close to the limits of stability, nuclear theory needs to become more predictable, because not all systems will be measured directly. For those nuclei that are studied experimentally, a deep understanding of the probe and its uncertainties is essential.  For these exotic systems, diverse reaction probes exist that enable the study of a wide variety of nuclear phenomena.  A solid understanding of reaction theory is crucial in the interpretation of these experiments.  This understanding must include the sources of uncertainty within the models.  

There are four main sources of uncertainty in reaction theory, as discussed in \cite{Lovell2015}.  These include approximations to (a) the few-body problem, (b) the effective interactions used, and (c) the structure functions (such as overlaps).  Most of these have been investigated, in \cite{Lovell2015} (and the references therein) and elsewhere.  However, these investigations typically rely on the comparison of two models or parameterizations.  Concerning (a), for example, methods such as the adiabatic approximation or continuum-discretized coupled channels method have been benchmarked against the Faddeev method \cite{Nunes2011,Deltuva2007,Capel2012}.  To address (b), the uncertainty from the effective interactions used, the standard procedure is to use two different parameterizations of the optical model within the same reaction theory framework, with the percent errors coming from the difference between the results obtained with these two parameterizations.  The same approach is taken when investigating (c), the effect of simplifications in the structure functions.  

Although comparative methods can be used to investigate each of these sources of uncertainties, they are not systematic. They also do not allow us to know, \emph{a priori}, when these effects will become important.  In order to move the field forward, systematic ways to compute uncertainties must be developed.  

As opposed to reaction theory, in other nuclear theory subfields, systematic methods of uncertainty quantification have become widespread.  Bayesian methods for parameter estimation are being used in effective field theories (EFTs), for example \cite{Wesolowski2016,Schindler2009}, as well as in nuclear data evaluations \cite{Koning2015}.  Truncation errors are being systematically investigated in EFTs and the derivation of the nuclear force \cite{Perez2015,Furnstahl2015,Perez2016,Carlsson2016}.  Uncertainty quantification has also been investigated in and applied to density functional theory (DFT), for example in \cite{McDonnell2015,Schunck2015,Schunck2015JPG}.

Uncertainty quantification has also been used widely in fields outside low energy nuclear physics.  For example, Bayesian approaches are used in measuring neutron star radii \cite{Ozel2015} and estimating parameters for heavy-ion collisions \cite{Bernhard2016}.  Of course, systematic error quantifications are also the topic of research in fields beyond physics.  The many lessons learned from the large array of applications can guide the work on uncertainty quantification in reaction theory; however, many specifics need to addressed to improve on the state of the art. Our approach therefore is to start with simple formulations of the process and focus on developing a systematic and robust methodology that holds specifically for nuclear reactions. This work represents the first step in this process.

Our goal here is to quantify uncertainties coming from parameter variations of the effective interaction within a given model.  In Section \ref{theory}, we discuss the formulation used for our study of uncertainties and summarize the reaction models that we are considering.  The systems that we have studied are introduced in Section \ref{cases}. Results are presented in Section \ref{results}, including a detailed example and a summary of the overall trends, and in Section \ref{discussion}, these results are discussed.  We conclude in Section \ref{conclusions} by presenting ideas for improving this analysis.  

%%%%%%%%%%%%%%%%%%%%%%%%%%%%%%%%%%%%%%%%%%%%%%%%%%%%%%%%%%%%%%%%%%%%%%%%%%%%%%%%%%%%%%%%%%%%%%%%%

\section{Theoretical framework}
\label{theory}

Throughout this work, we assume that all uncertainties are coming from our choice of parameterization of the interaction and not from the reaction model itself.  Uncertainties in the reaction model are discussed briefly in Section \ref{conclusions}.  

\subsection{Uncorrelated $\chi^2$ fitting}
\label{uncorr}

%\begin{itemize}
%\item Assumptions in the model (what is normally distributed)
%\item Definition of the $\chi^2$ fitting function
%\item Process to define 95\% confidence bands
%\end{itemize}

The goal is to describe a true function, $\sigma(\theta)$, with a known model, $\sigma^{\mathrm{th}}(\textbf{x},\theta)$, where the model is a function of angle, $\theta$, and has $N$ free parameters $\textbf{x} = (x_1,...,x_N)$.  In this work, the model is a formulation for a differential cross section, and the free parameters are the parameters in the optical potential; these are discussed in Section \ref{xsmodels}.  The model describes $M$ sets of data points $\{(\theta _1, \sigma^{\mathrm{exp}}_1),...,(\theta _M,\sigma^{\mathrm{exp}}_M)\}$, each with an associated experimental error, $\Delta \sigma_i$.  Typically, the data and errors are independent of one another and normally distributed, such that
\begin{equation}
\sigma^{\mathrm{exp}}_i = \sigma(\theta_i) + \epsilon _i,
\end{equation}

\noindent with the measurement errors being described as
\begin{equation}
\epsilon _i \sim \mathcal{N}(0,(\Delta \sigma _i) ^2),
\end{equation}

\noindent where $\mathcal{N}$ is the normal distribution.

In matrix form, this is
\begin{equation}
\label{eqn:expdist}
\sigma^{\mathrm{exp}} \sim \mathcal{N}(\sigma, \Sigma),
\end{equation}

\noindent where $\Sigma$ is an $M \times M$ diagonal matrix with $(\Delta \sigma _i) ^2$ on the diagonal.

The residuals, the differences between the data and the model, then have the multivariate normal distribution:
\begin{equation}
\label{eqn:uncorres}
\left [\sigma^{\mathrm{th}}(\textbf{x},\theta _1)-\sigma^{\mathrm{exp}}_1,..., \sigma^{\mathrm{th}}(\textbf{x},\theta_M)-\sigma^{\mathrm{exp}}_M \right ]^T \sim \mathcal{N}(0,\Sigma).
\end{equation}

\noindent Maximizing the associated likelihood of \textbf{x} gives us the minimization objective function,
\begin{equation}
\label{eqn:uncorrchi}
\chi ^2 _{UC} (\textbf{x}) = \sum \limits _{i=1} ^M \left ( \frac{\sigma^{\mathrm{th}}(\textbf{x},\theta_i) - \sigma^{\mathrm{exp}}_i}{\Delta \sigma _i} \right )^2,
\end{equation}

\noindent which is proportional to the definition of the standard $\chi^2$ function (here, $UC$ stands for uncorrelated).  In minimizing this function, we can find the best-fit set of parameters, $\hat{\textbf{x}}$.

From here, we can define the $95\%$ confidence region about $\hat{\textbf{x}}$.  To do so, we further assume that the true parameter values are normally distributed around the minimum of the $\chi^2$ function, 
\begin{equation}
\label{eqn:MVG}
\mathcal{N}(\hat{\textbf{x}},\mathbb{C}_p) \sim \mathrm{exp}[-\frac{1}{2}(\textbf{x}-\hat{\textbf{x}})^T\mathbb{C}_p(\textbf{x}-\hat{\textbf{x}})],
\end{equation}

\noindent where $\mathbb{C}_p$ is the $N \times N$ parameter covariance matrix, describing the correlations between the fit parameters \cite{Regression}.  This assumption of a normal distribution can be supported empirically by looking at the two-dimensional slices of parameter space as a function of $\chi^2_{UC}$ value.  To take into account the goodness of the fit, we scale the parameter covariance matrix by the degrees of freedom
\begin{equation}
\label{eqn:ssq}
s^2 = \frac{\chi^2_{UC}}{M-N},
\end{equation}

\noindent such that in Eq. \ref{eqn:MVG} $\mathbb{C}_p$ is replaced by $s^2 \mathbb{C}_p$.

Parameter sets can then be drawn from the scaled distribution (\ref{eqn:MVG}) and run through the model $\sigma ^{\mathrm{th}}$.  At each angle where the model was evaluated, the highest 2.5\% and lowest 2.5\% of the calculations are removed in order to obtain a 95\% confidence band.  

We can then define the average width of the uncorrelated confidence band as
\begin{equation}
\label{eqn:avgwidth}
\overline{W}_{UC} = \frac{1}{N_\theta} \sum \limits _{i=1}^{N_\theta} (\sigma ^{\mathrm{max}}_i - \sigma^{\mathrm{min}}_i),
\end{equation}

where $\sigma^{\mathrm{max}}_i$ ($\sigma^{\mathrm{min}}_i$) is the cross section value at the upper (lower) limit of the 95\% confidence band for a given angle $\theta_i$ and $N_\theta$ is the number of angles included in the calculation.

\subsection{Correlated $\chi^2$ fitting}
\label{corr}

%\begin{itemize}
%\item Reason for needing a correlated fitting function
%\item Extra assumptions that are now added
%\item Definition of the new $\chi^2$ fitting function
%\item From here, confidence bands are defined in the same way as previous section
%\end{itemize}

If there are correlations between the $\epsilon _i$ errors (e.g., between the predicted values of the cross section for different angles), one needs to take a different approach.  

Consider, for example, the single-channel elastic cross section, typically expressed by a partial wave decomposition as
\begin{equation}
\frac{d\sigma}{d\Omega} = \frac{1}{4k^2} \left | \sum \limits _{L=0} ^\infty (2L+1)P_L(\mathrm{cos}\theta)(S_L -1)\right |^2,
\end{equation}

\noindent where $k$ is the incoming momentum, $P_L(\mathrm{cos}\theta)$ are the Legendre polynomials, and $S_L$ is the corresponding S-matrix \cite{ReactionsBook}.  Because fitting at one angle influences the cross-section values at all other angles, the correlations between angles may need to be taken into account in the fitting process.

Therefore, along with the condition of Eq. \ref{eqn:expdist}, we assume
\begin{equation}
[\sigma^{\mathrm{th}}(\textbf{x},\theta_1),...,\sigma^{\mathrm{th}}(\textbf{x},\theta_M)]^T \sim \mathcal{N}(\sigma,\mathbb{C}_m);
\end{equation}

\noindent that is, our model is normally distributed with the $M \times M$ model covariance matrix, $\mathbb{C}_m$, which describes the correlations within the model at each of the experimentally measured angles.  This leaves the residuals distributed as 
\begin{equation}
\left [\sigma^{\mathrm{th}}(\textbf{x},\theta _1)-\sigma^{\mathrm{exp}}_1,...,\sigma^{\mathrm{th}}(\textbf{x},\theta_M)-\sigma^{\mathrm{exp}}_M \right ]^T \sim \mathcal{N}(0,\mathbb{C}_m+\Sigma).
\end{equation}

The resulting correlated $\chi ^2$ minimization objective function becomes
\begin{equation}
\label{eqn:corrchi}
\chi^2 _C (\hat{\textbf{x}}) = \sum \limits _{i=1}^M \sum \limits _{j=1}^M W_{ij}(\sigma^{\mathrm{th}}(\textbf{x},\theta_i)-\sigma^{\mathrm{exp}}_i)(\sigma^{\mathrm{th}}(\textbf{x},\theta_j)-\sigma^{\mathrm{exp}}_j),
\end{equation}

\noindent where $W_{ij}$ are the $(ij)^{\mathrm{th}}$ matrix elements of $\mathbb{W} = (\mathbb{C}_m + \Sigma)^{-1}$.  Note that interference can occur between the residuals at different angles because the individual model covariance matrix elements $W_{ij}$ do not have to be positive.  This causes $\chi^2_{C}$ to be less than $\chi^2_{UC}$.  In general, we see that $\chi^2_{UC}/M\le1$ is no longer the definition of a statistical fit because the model covariance matrix is not normalized.  

Once the set of parameters corresponding to the best fit is found with this minimization objective function, 95\% confidence bands can be defined in the same way as described as in Section \ref{uncorr}, substituting $\chi^2_{C}$ for $\chi^2_{UC}$.  The average width of the correlated confidence bands can be defined identically to Eq. \ref{eqn:avgwidth} where $\sigma^{\mathrm{max}}_i$ and $\sigma^{\mathrm{min}}_i$ are associated with the correlated 95\% confidence bands.

\subsection{Matrices}

%\begin{itemize}
%\item Define the model covariance matrix - since this is a little different?
%\item Define Jacobian, covariance and correlation matrices
%\end{itemize}

The parameter covariance matrix is defined as 
\begin{equation}
\mathbb{C}_p = (\mathbb{J}^T \mathbb{J})^{-1},
\end{equation}

\noindent where the matrix elements of the Jacobian, $\mathbb{J}$, are
\begin{equation}
\mathbb{J}_{ij} = \left. \frac{\partial \sigma^{\mathrm{th}}(\textbf{x},\theta_i)}{\partial x_j} \right |_{\textbf{x} = \hat{\textbf{x}}}.
\end{equation}

From the covariance matrix, the parameter correlation matrix can be defined \cite{Regression} as
\begin{equation}
\mathbb{C}_{corr} = \mathbb{A}^T \mathbb{C}_p \mathbb{A}.
\end{equation}

\noindent Here, $\mathbb{A}$ is the matrix that has, as its diagonal elements, the inverse of the square roots of the diagonal elements of $\mathbb{C}_p$ and zeros on its off-diagonal elements ($\mathbb{A}_{ii} = 1/\sqrt{(\mathbb{C}_p)_{ii}}$).  The magnitude of the matrix elements of the correlation matrix range from zero to one.  Zero means no correlation between the two parameters, and (negative) one means that the two parameters are fully (anti-)correlated.  Therefore, the diagonal elements of the correlation matrix are all one, since every parameter is fully correlated with itself.  

\subsection{Non-Gaussian parameter space}

%\begin{itemize}
%\item pulling from the $\chi^2$ distribution if parameter space is not Gaussian
%\item Check your equations
%\end{itemize}

The $\chi^2$ function around the minimum in parameter space may not be approximately quadratic, an assumption required by Eq. \ref{eqn:MVG}.  If this is the case, we can pull the parameter sets from the actual distribution (defined by contours of constant $\chi^2$), rather than the multivariate Gaussian of Eq. \ref{eqn:MVG}.  For each parameter set pulled, its $\chi^2(\textbf{x})$ value is tested in the inequality
\begin{equation}
\label{eqn:nongausparm}
\chi^2(\textbf{x}) - \chi^2(\hat{\textbf{x}}) \le 9M,
\end{equation}

\noindent which gives sets of parameters that are approximately within $3$ standard deviations of the minimum \cite{ReactionsBook}.

If the $\chi^2(\textbf{x})$ associated with the parameter set fulfills the requirement of Eq. \ref{eqn:nongausparm}, the parameter set is kept; otherwise, it is thrown away.  When 200 parameter sets have been accepted, these are run through the model.  Since Eq. \ref{eqn:nongausparm} defines a 99.7\% region, confidence bands must be slightly expanded to represent the 0.3\% of parameters sets that are rejected.  This is accounted for by removing the highest 2.35\% and lowest 2.35\% of the calculations at each angle.

\subsection{Reaction models}
\label{xsmodels}

%\begin{itemize}
%\item Small section about DWBA and CCBA
%\end{itemize}

In this work, we fit elastic-scattering data to predict inelastic-scattering cross sections as well as transfer cross sections.  Because the focus of this work is to introduce methodology to systematically quantify uncertainties in predictions of nuclear reactions, here, we try to keep the reaction mechanisms as simple as possible.  Therefore, we focus on two reaction models, the coupled-channels Born approximation (CCBA) when performing elastic- and inelastic-scattering calculations and the distorted-wave Born approximation (DWBA) when performing elastic scattering and transfer calculations \cite{ReactionsBook}.  Improvements on reaction models themselves will be performed elsewhere at a later stage.

\subsubsection{CCBA}

CCBA couples the elastic and inelastic channels together by solving $N$ coupled-channel equations,
\begin{equation}
[H_{\alpha}-E_{\alpha}]\psi _{\alpha \alpha_{i}}(R_\alpha) + \sum \limits _{\beta \ne \alpha} ^N V_{\alpha \beta} \psi _{\beta \alpha_{i}} (R_\beta) = 0,
\end{equation}

\noindent where $\alpha _i$ denotes the incoming elastic channel, $\psi_{\alpha \alpha_{i}}$ is the two-body wave function for a given outgoing channel $\alpha$, and $V_{\alpha \beta}$ is the coupling potential \cite{ReactionsBook}.

In this work, CCBA is used when calculating neutron elastic- and inelastic-scattering cross sections, coupling only the first excited state to the ground state.

\subsubsection{DWBA}

For $A(d,p)B$ reactions, the one-step distorted wave approximation is made.  The elastic scattering of the deuteron is described by an effective deuteron-target interaction, $V_{dA}$.  For the transfer reaction, in DWBA, instead of solving the true scattering three-body $d+A$ problem, the three-body deuteron scattering wave function is replaced by the deuteron elastic component, namely a deuteron distorted wave multiplied by the corresponding bound state of the deuteron \cite{ReactionsBook}:
\begin{equation}
\textbf{T}^{\mathrm{DWBA}}_{\mathrm{post}} = \langle \Phi_{nA} (\vec{r}_{nA}) \chi_p (\vec{R}_f) | V_{np}+\Delta| \Phi _{np}(\vec{r}_{np}) \chi _{d\vec{k}_{i}}(\vec{R}_i) \rangle.
\end{equation}

\noindent Here, $\Phi_{np} (\vec{r}_{np})$ is the initial bound-state wave function for the deuteron, $\chi _{d\vec{k}_{i}}(\vec{R}_i)$ is the distorted wave of the $d+A$ system, $\Phi_{nA} (\vec{r}_{nA})$ is the final bound-state wave function of $B$, $\chi _p(\vec{R}_f)$ is the distorted wave of the outgoing proton interacting with $B$, $V_{np}$ is the deuteron binding potential, and $\Delta$ is the difference between the $A+p$ and $B+p$ optical potentials.

After calculating the transfer cross section, spectroscopic factors are typically extracted from
\begin{equation}
\left ( \frac{d\sigma}{d\Omega} \right ) ^{\mathrm{exp}} = S^{\mathrm{exp}} \left ( \frac{d\sigma}{d\Omega} \right) ^{\mathrm{DWBA}}
\end{equation}

\noindent at the first peak of the angular distributions.  $S^{\mathrm{exp}}$ describes the single-particle nature of the transfered nucleon in the composite nucleus, $B$, relative to nucleus $A$ \cite{ReactionsBook}.

\subsubsection{Optical model}

For both the CCBA and DWBA calculations, optical potentials are used.  These are characterized by a real and an imaginary part,
\begin{equation}
U(r) = V(r) + iW(r).
\end{equation}

\noindent The imaginary part takes into account the flux that leaves the elastic channel and is not explicitly described by the model.

These potentials have volume, surface, and spin-orbit parts that are characterized by a Woods-Saxon shape or derivatives of a Woods-Saxon shape.  If we consider
\begin{equation}
\label{eqn:real}
V(r) = -\frac{V_o}{1+exp(\frac{r-R_o}{a_o})}
\end{equation}

\noindent and
\begin{equation}
\label{eqn:imag}
W(r) = -\frac{W_V}{1+exp(\frac{r-R_w}{a_w})}
\end{equation}

\noindent for the volume terms, there are six free parameters in the fit.  In this parameterization, $R_i = r_i A^{1/3}$, where $A$ is mass number and the fitted parameter is $r_i$.  The surface term is defined by the derivative of a Woods-Saxon shape and is typically purely imaginary, which introduces another three parameters, $W_s$, $r_s$, and $a_s$.  The spin-orbit potential is also parameterized by the derivative of a Woods-Saxon shape; however, to limit the number of free parameters for this introductory work, we keep all the spin-orbit parameters fixed at the original values from the parameterizations referenced in Table \ref{tab:cases}.  The Coulomb potential is included in the usual way (e.g., \cite{Fukui2014}) and is parameterized by a single Coulomb radius, which is also kept fixed throughout this work.

%%%%%%%%%%%%%%%%%%%%%%%%%%%%%%%%%%%%%%%%%%%%%%%%%%%%%%%%%%%%%%%%%%%%%%%%%%%%%%%%%%%%%%%%%%%%%%%%%
\section{Numerical Details}
\label{cases}

%\begin{itemize}
%\item For all cases, fit elastic to predict transfer or inelastic
%\item Two transfer cases - $^{12}$C and $^{90}$Zr
%\item Four inelastic cases - $^{12}$C, $^{48}$Ca, $^{54}$Fe, and $^{208}$Pb
%\item Table with cases, references for data and starting optical model fit parameters
%\end{itemize}

In each of the cases studied, elastic-scattering data was fit, and then either inelastic-scattering or transfer cross sections were predicted and compared to data.  Table \ref{tab:cases} summarizes the systems that were studied, including references to the starting optical model and corresponding data set.

\begin{table}[h]
\begin{center}
\begin{tabular}{| c | c | c | c |}
\hline \textbf{System} & \textbf{Energy (MeV)} & \textbf{Data} & \textbf{Potential} \\ \hline
$^{12}$C(d,d)$^{12}$C & 11.8 & \cite{12Cd} & \cite{12Cdpot} \\ \hline
$^{12}$C(d,p)$^{13}$C & 11.8 & \cite{12Cdp} & --- \\ \hline
$^{90}$Zr(d,d)$^{90}$Zr & 12.0 & \cite{90Zrdd} & \cite{90Zrdpot} \\ \hline
$^{90}$Zr(d,p)$^{91}$Zr & 12.0 & \cite{90Zrdp} & --- \\ \hline
$^{12}$C(n,n)$^{12}$C & 17.29 & \cite{12Cnn} & \cite{12Cnnpot} \\ \hline
$^{12}$C(n,n')$^{12}$C(2$^+_1$) & 17.29 & \cite{12Cnn} & --- \\ \hline
$^{48}$Ca(n,n)$^{48}$Ca & 7.97 & \cite{48Cann} & \cite{48Canpot} \\ \hline
$^{48}$Ca(n,n')$^{48}$Ca(2$^+_1$) & 7.97 & \cite{48Cann} & --- \\ \hline
$^{54}$Fe(n,n)$^{54}$Fe & 16.93 & \cite{54Fenn} & \cite{54Fenpot} \\ \hline
$^{54}$Fe(n,n')$^{54}$Fe(2$^+_1$) & 16.93 & \cite{54Fenn} & --- \\ \hline
$^{208}$Pb(n,n)$^{208}$Pb & 26.0 & \cite{208Pbnel} & \cite{54Fenn} \\ \hline
$^{208}$Pb(n,n')$^{208}$Pb(3$^-_1$) & 26.0 & \cite{208Pbninel} & --- \\ \hline
\end{tabular}
\end{center}
\caption{Reactions studied in this work.  The third column gives the reference for the corresponding data set, and the fourth column gives the reference for the starting optical model parameterization.}
\label{tab:cases}
\end{table}

The $\delta_2$ values used in this work for the inelastic scattering of $^{12}$C, $^{48}$Ca, and $^{54}$Fe are 1.0852 fm, 0.85 fm, and 0.967 fm, respectively.  For $^{208}$Pb, $\delta_3 = 0.296$ fm for the uncorrelated fit, and $\delta_3 = 0.230$ fm for the correlated fit.  This difference was introduced to better match the magnitude of the calculated inelastic-scattering cross sections to the data.  All values were adjusted from \cite{nndc} to better describe the magnitude of the inelastic cross sections.

For the two $(d,p)$ reactions, the outgoing channels were defined by the potentials described in \cite{12Cdp} for the $^{12}$C(d,p) and \cite{90Zrdp} for the $^{90}$Zr(d,p) reactions.  The binding potential between the target and transfered neutron was described by a Woods-Saxon shape with radius of 1.2 fm and diffuseness of 0.60 fm.  The depth of this potential was adjusted to reproduce the experimental binding energy.  A spin-orbit potential was also included, with standard depth, radius, and diffuseness of 7.0 MeV, 1.2 fm, and 0.60 fm, respectively.  The $np$ interaction for the deuteron is taken from \cite{Reid1968}.

The statistical approach described in Section \ref{theory} is newly implemented, but it makes use of the reaction codes FRESCO and SFRESCO (which employs the MINUIT \cite{minuit} minimization routines) \cite{fresco}.

%%%%%%%%%%%%%%%%%%%%%%%%%%%%%%%%%%%%%%%%%%%%%%%%%%%%%%%%%%%%%%%%%%%%%%%%%%%%%%%%%%%%%%%%%%%%%%%%%

\section{Results}
\label{results}

%\begin{itemize}
%\item Uncorrelated fit - potential values, 2D contours, 95\% CB, covariance matrix, why we need a correlated fit
%\item Correlated fit - potential values, comparisons of fits, 2D contours, 95\% CB, covariance matrix
%\item Fitting one data set (first two bullets)
%\item Fitting both data sets (first two bullets)
%\item Pick a case to show explicitly - $^{54}$Fe(n,n)$^{54}$Fe and $^{54}$Fe(n,n')$^{54}$Fe(2$^+_1$)
%\item Put in $\chi^2$ values for each fit (in the detailed section)
%\item Put in confidence band widths for each fit (in the detailed section)
%\item Information from summary tables - average width of bands, $\chi^2$ values, goodness of fit/parameters
%\item Residual plots (elastic and inelastic)
%\item Model correlation matrix visuals (elastic and inelastic)
%\end{itemize}

To demonstrate our method, we will consider one of the cases from Table \ref{tab:cases} in detail:  fitting $^{12}$C(d,d)$^{12}$C elastic scattering to predict the $^{12}$C(d,p)$^{13}$C transfer cross section.  We focus on the difference between uncorrelated and correlated fitting.  The results from all other reactions are then summarized.

\subsection{Detailed example}

Starting from the parameterization referenced in Table \ref{tab:cases} for $^{12}$C(d,d)$^{12}$C and using the $\chi^2_{UC}$ minimization function of Eq. \ref{eqn:uncorrchi}, we reach the best-fit parameterization of Table \ref{tab:uncorrfit} (UC).  The blue parameters were simultaneously minimized, and their variations give rise to the 95\% confidence bands; the green parameters were initially varied but fixed during the final fitting procedure in order to keep these parameter values from becoming unphysical during the minimization process.  

\begin{table}[h]
\begin{center}
\begin{tabular}{| c | c | c | c | c | c | c |}
\hline & \textbf{V$_\mathrm{o}$(MeV)} & \textbf{r$_\mathrm{o}$(fm)} & \textbf{a$_\mathrm{o}$(fm)} & \textbf{W$_\mathrm{s}$(MeV)} &  \textbf{r$_\mathrm{s}$(fm)} & \textbf{a$_\mathrm{s}$(fm)} \\ \hline
UC & \IG{111.505} & \IB{1.002} & \IB{0.7308} & \IB{27.582} & \IB{1.235} & \IB{0.2841} \\ \hline
C & \IB{55.126} & \IB{1.121} & \IB{0.6700} & \IG{40.931} & \IB{1.193} & \IB{0.1963} \\ \hline
\end{tabular}
\end{center}
\caption{(Color online)  Best-fit parameters when $^{12}$C(d,d)$^{12}$C cross sections were fit to predict the $^{12}$C(d,p)$^{13}$C transfer cross section.  The second row gives the parameters for the uncorrelated fit (UC), and the third row gives the parameters for the correlated fit (C).  The second, third, and fourth (fifth, sixth, and seventh) columns are the real volume (imaginary surface) potential parameters.  Blue parameters were simultaneously minimized, while the green parameters were fixed during the final fitting procedure.}
\label{tab:uncorrfit}
\end{table}

\begin{figure}
\begin{center}
\begin{tabular}{cc}
\includegraphics[width=0.23\textwidth]{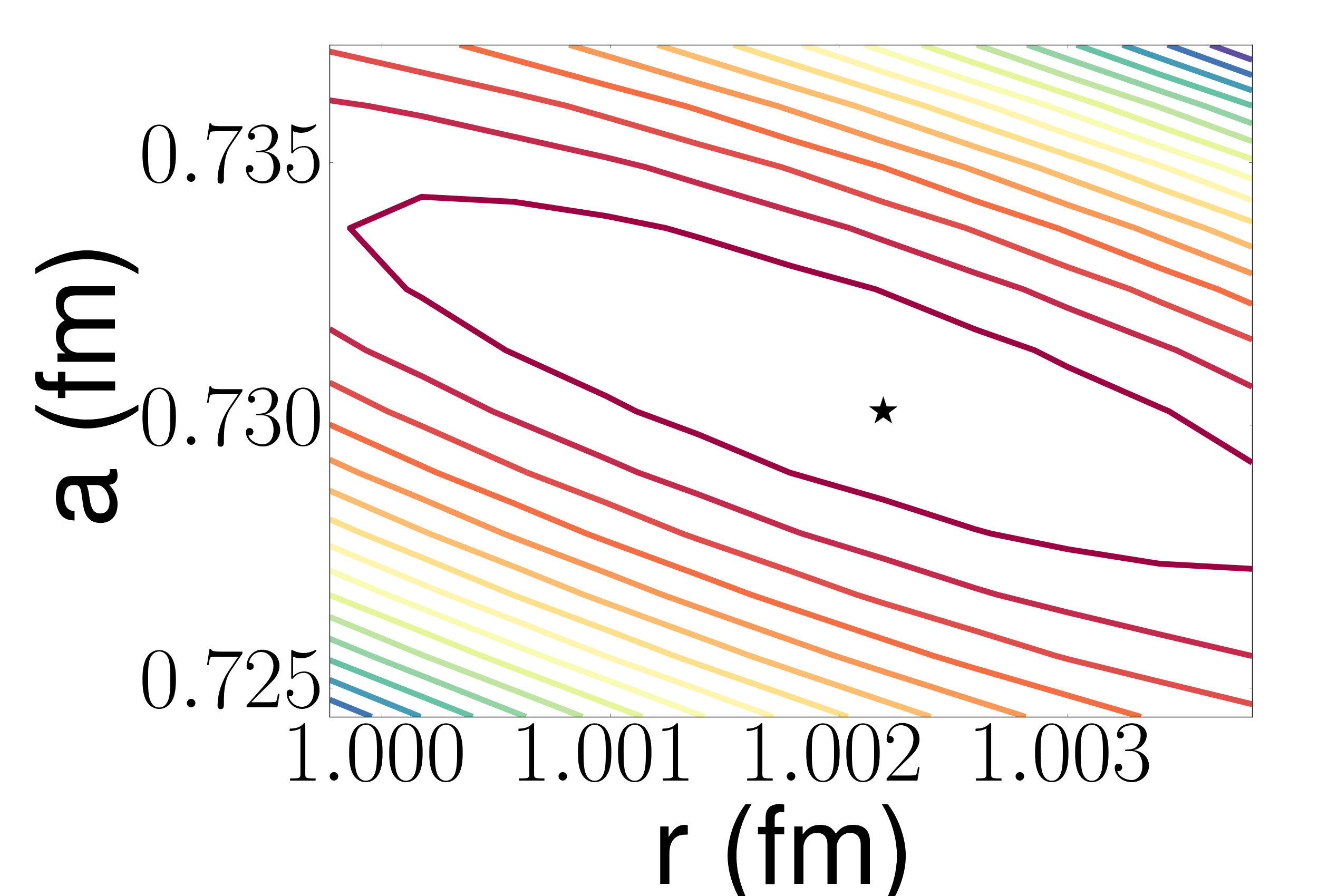} & \includegraphics[width=0.23\textwidth]{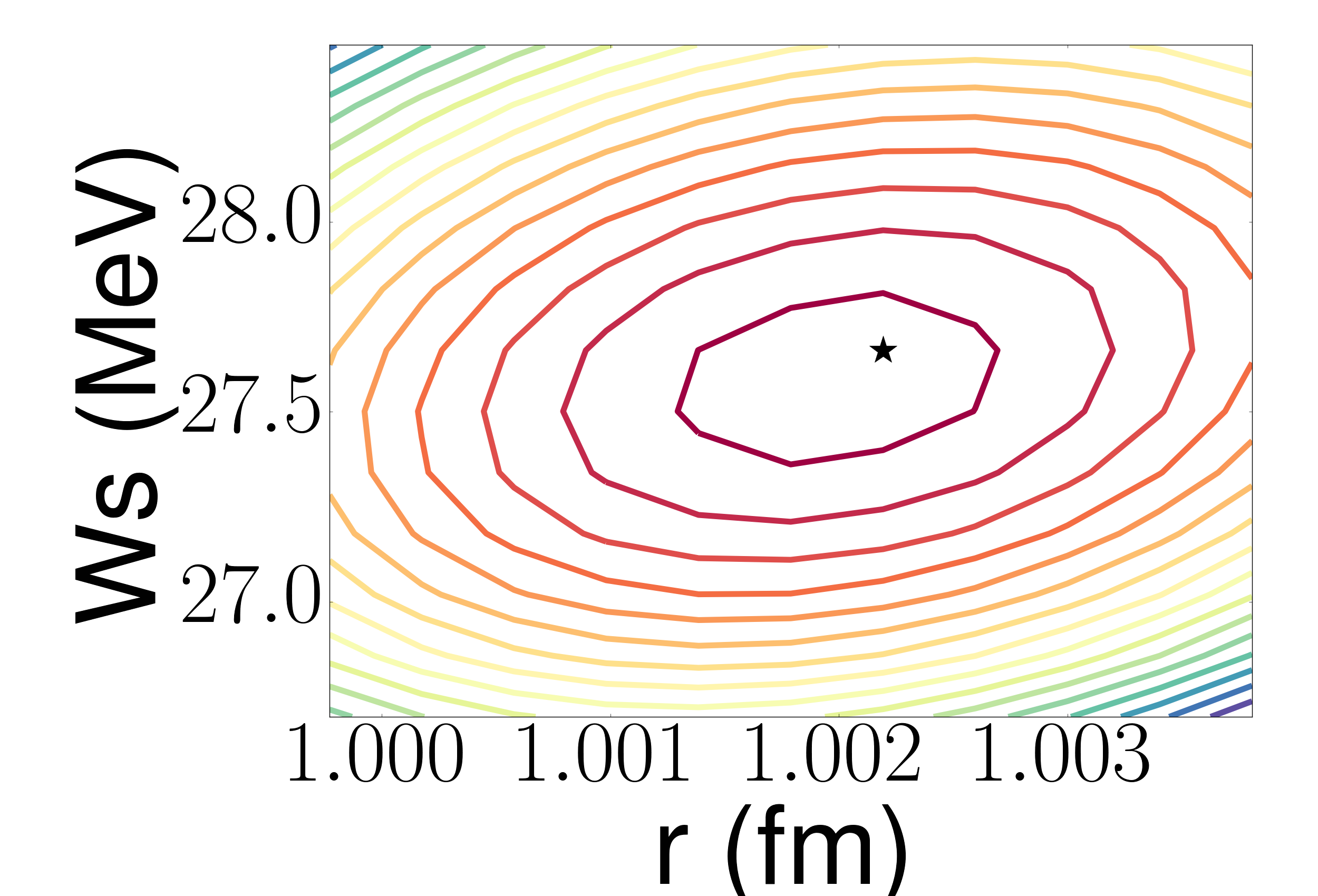} \\ 
\includegraphics[width=0.23\textwidth]{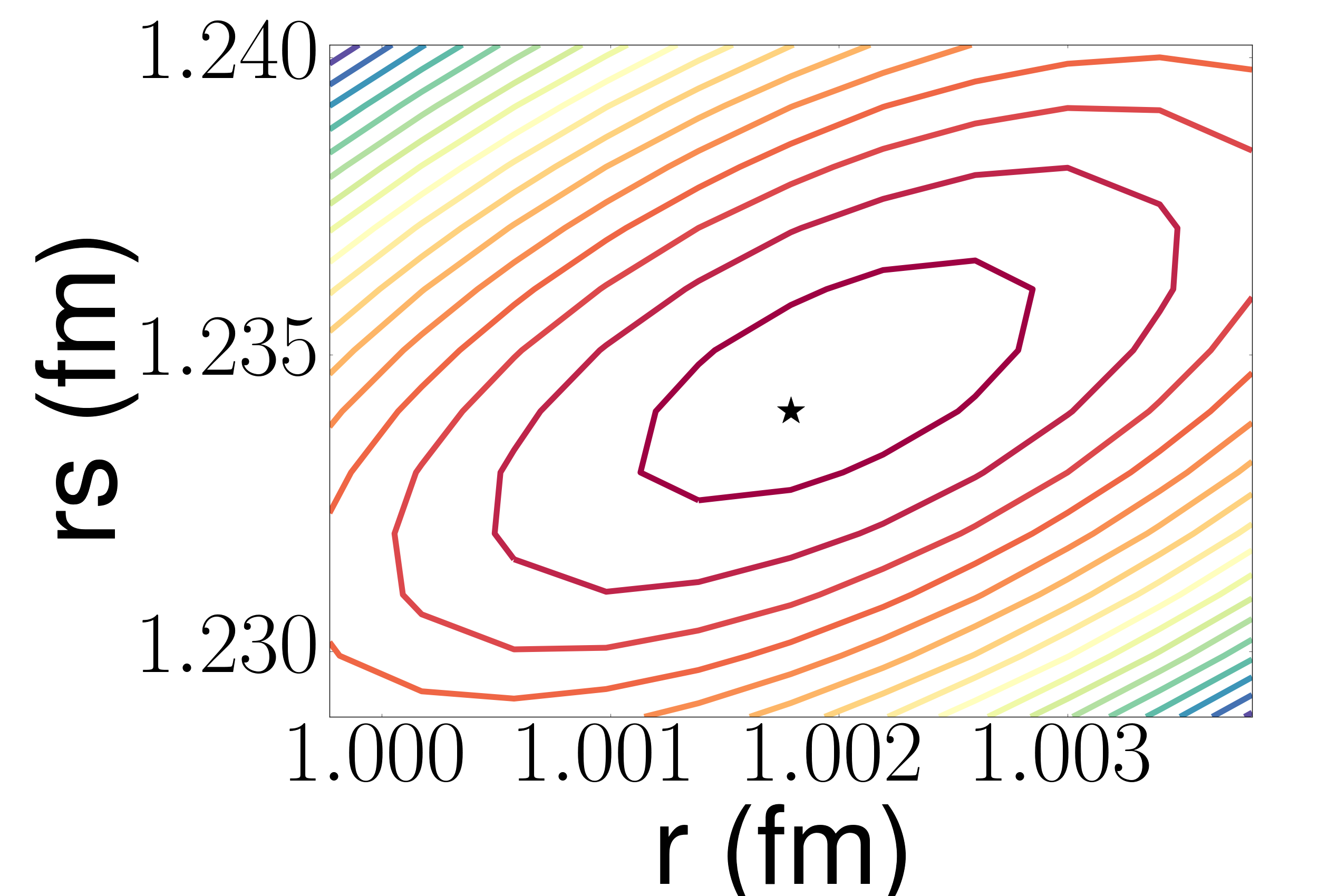} & \includegraphics[width=0.23\textwidth]{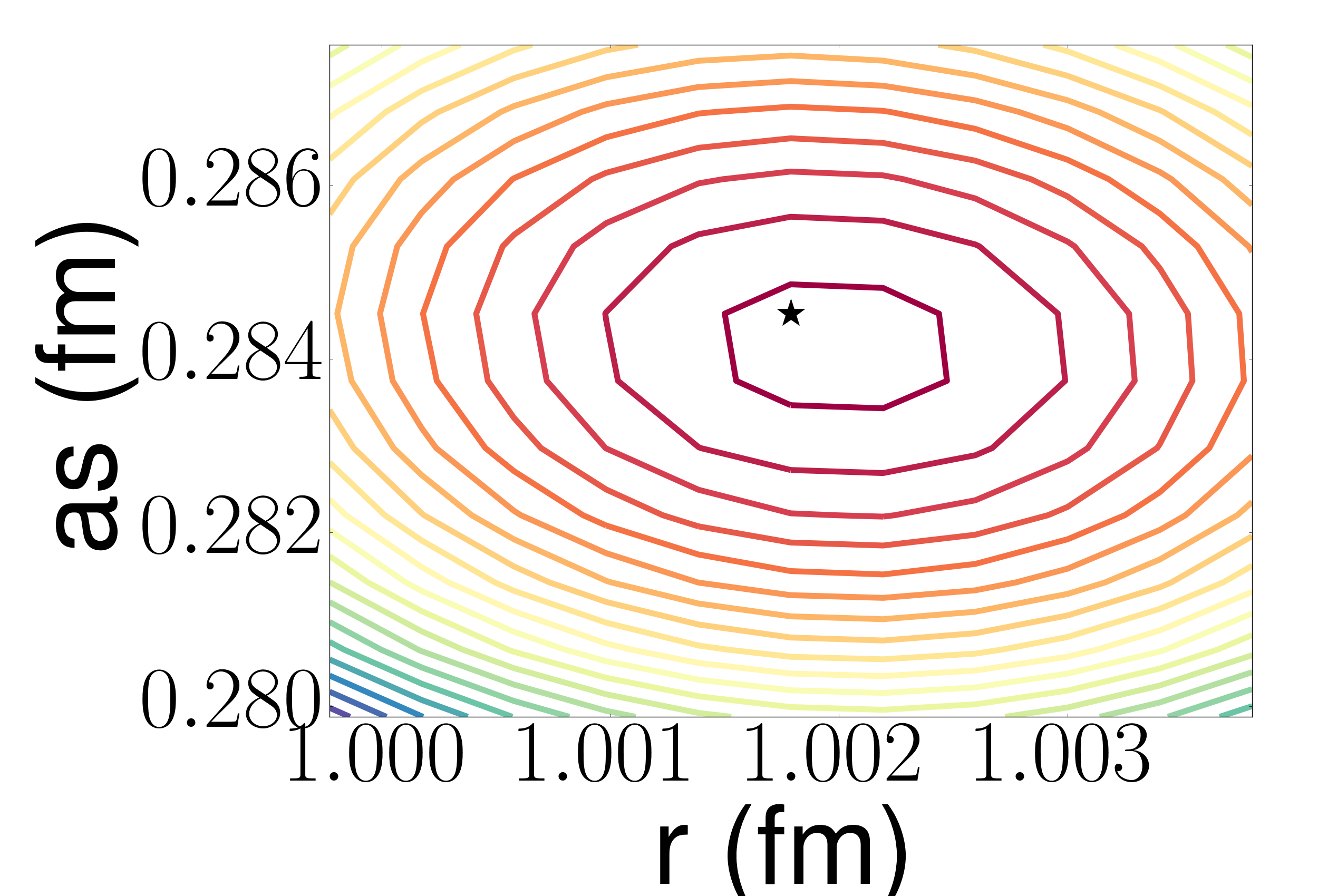} \\
\includegraphics[width=0.23\textwidth]{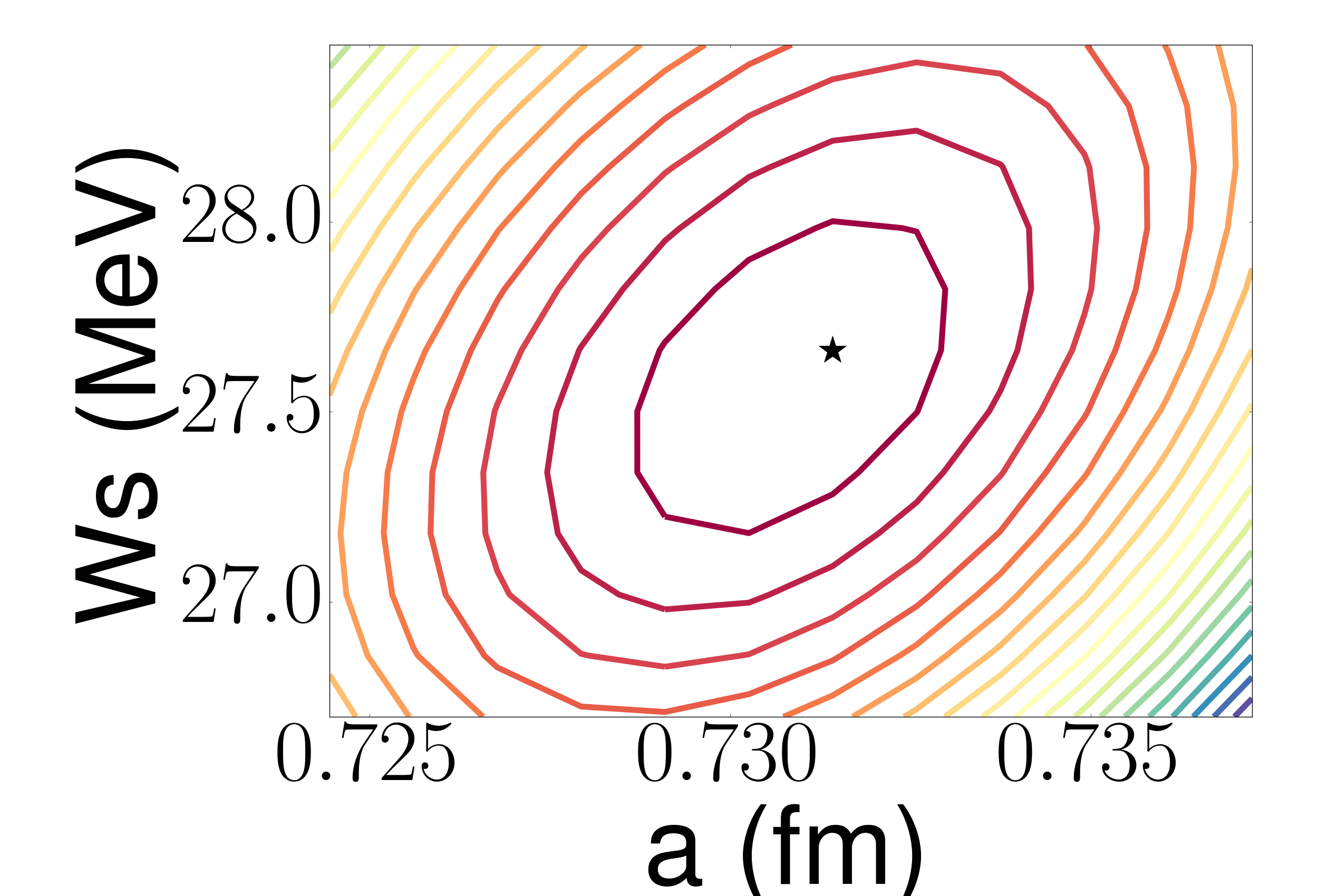} & \includegraphics[width=0.23\textwidth]{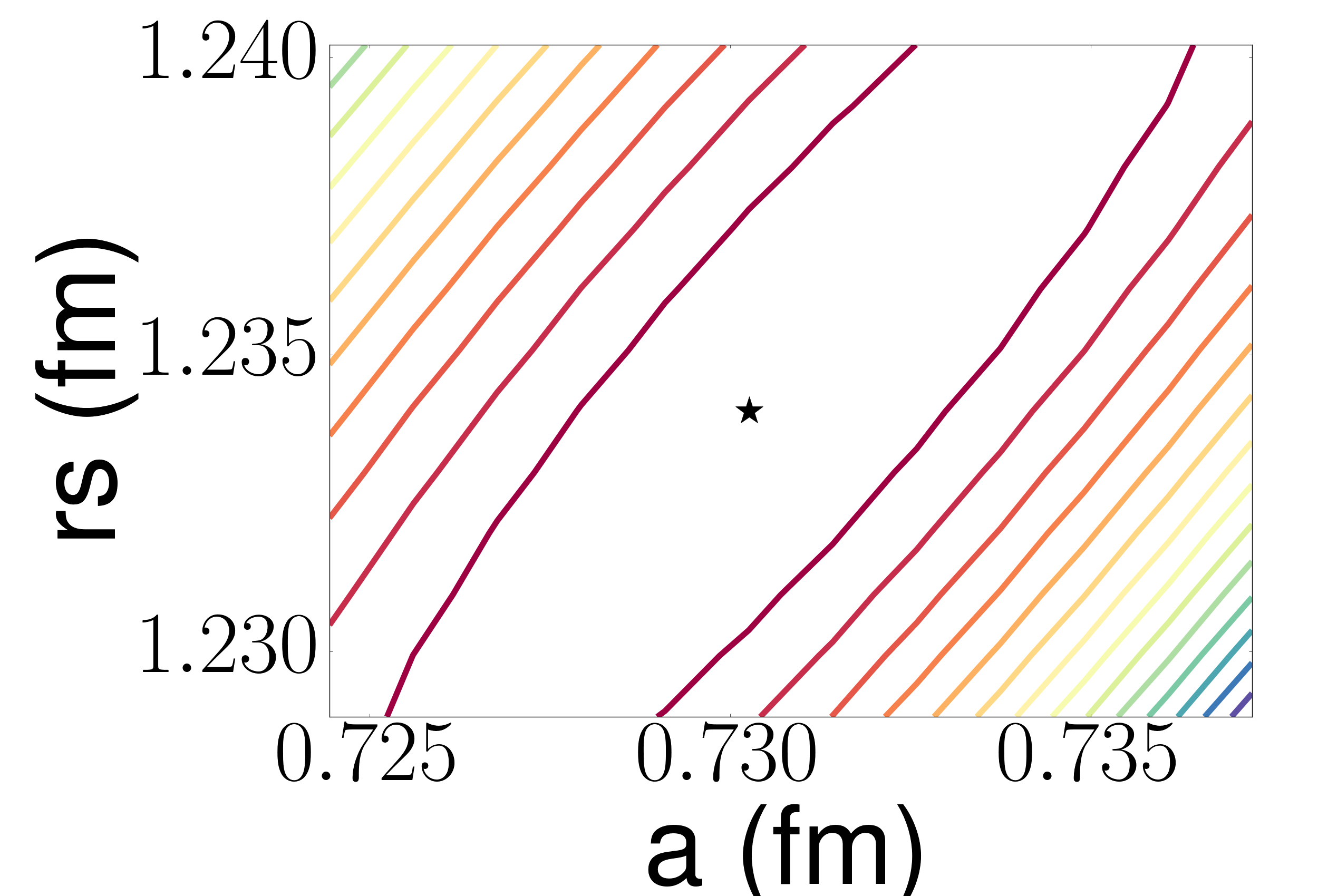} \\
\includegraphics[width=0.23\textwidth]{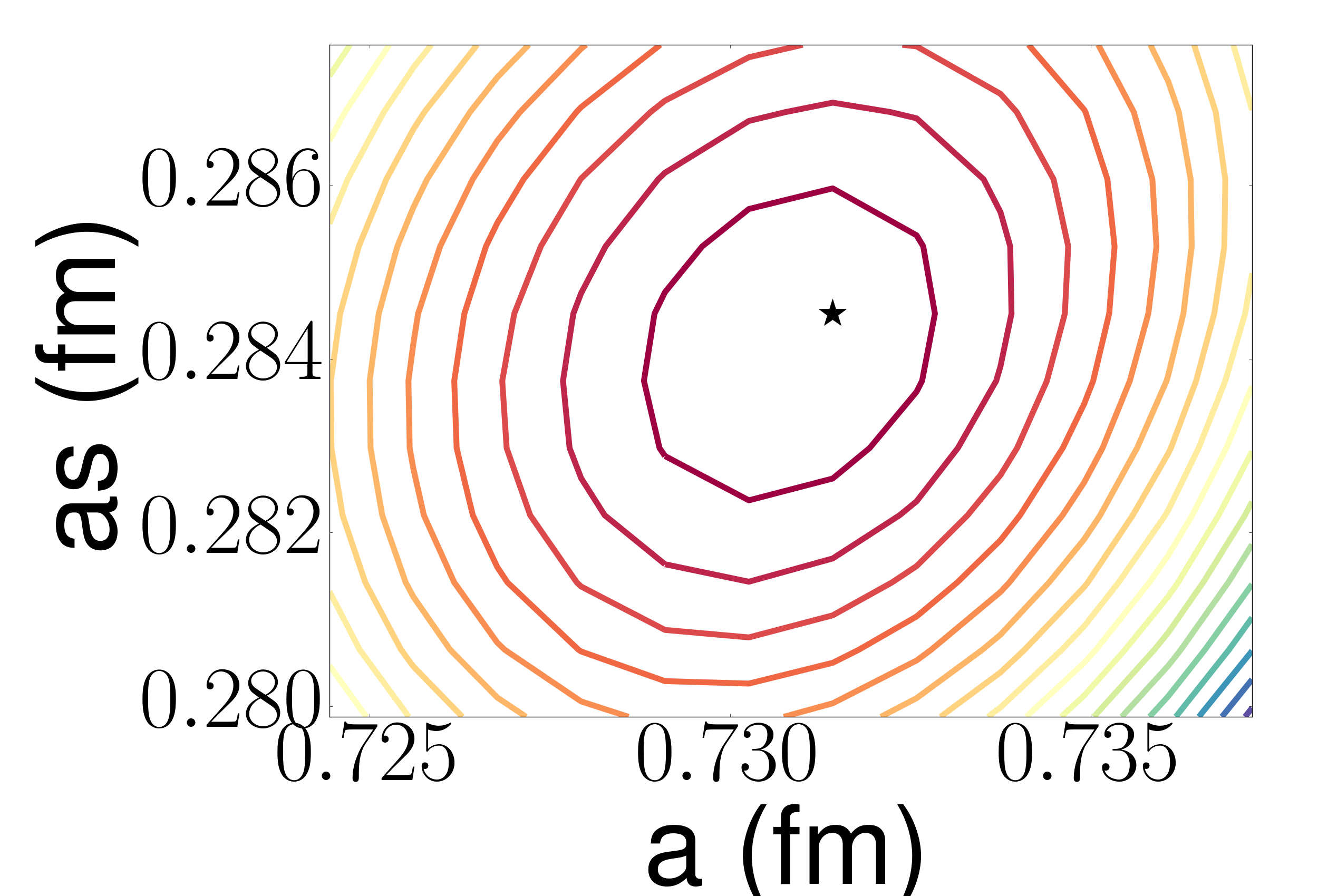} & \includegraphics[width=0.23\textwidth]{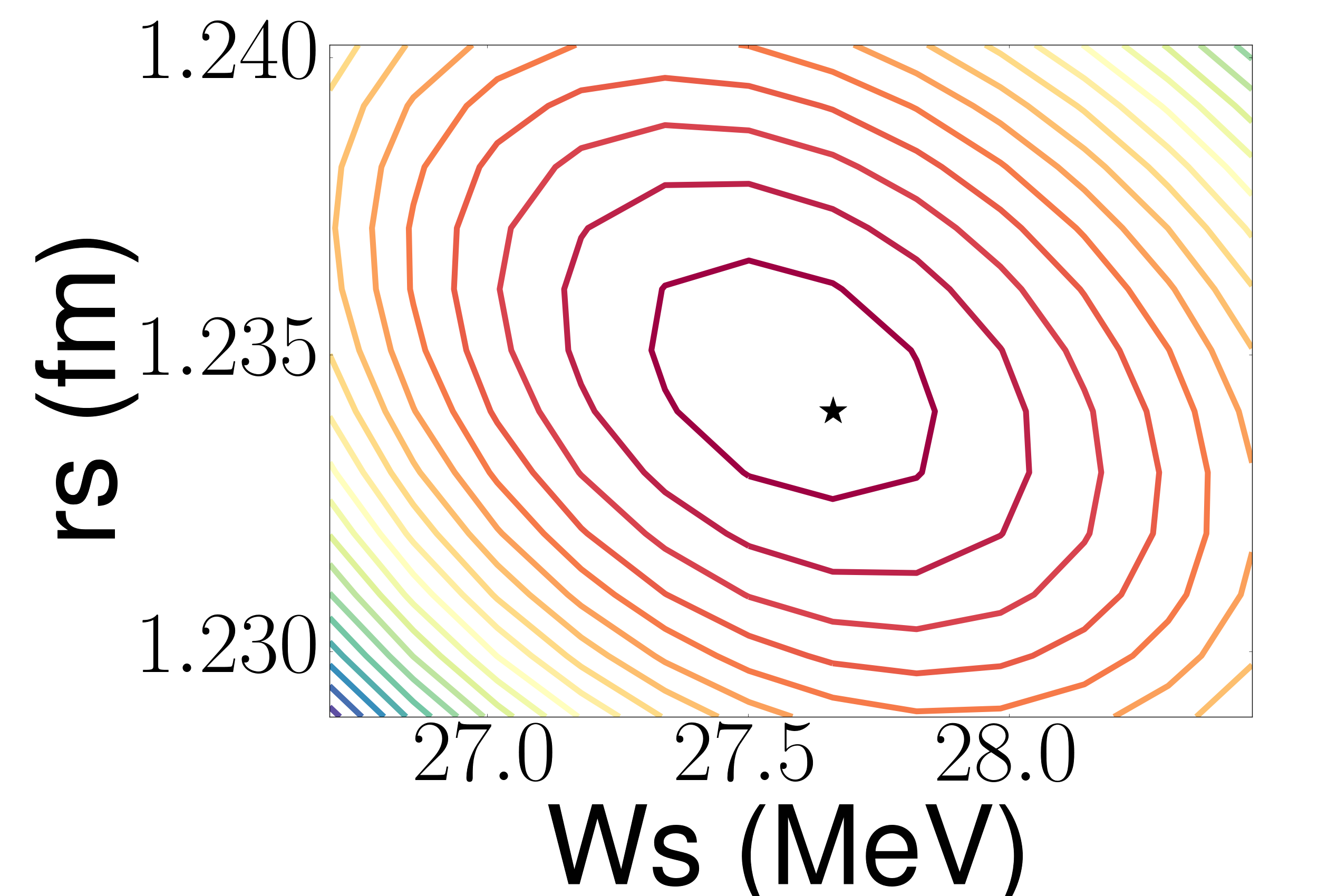} \\
\includegraphics[width=0.23\textwidth]{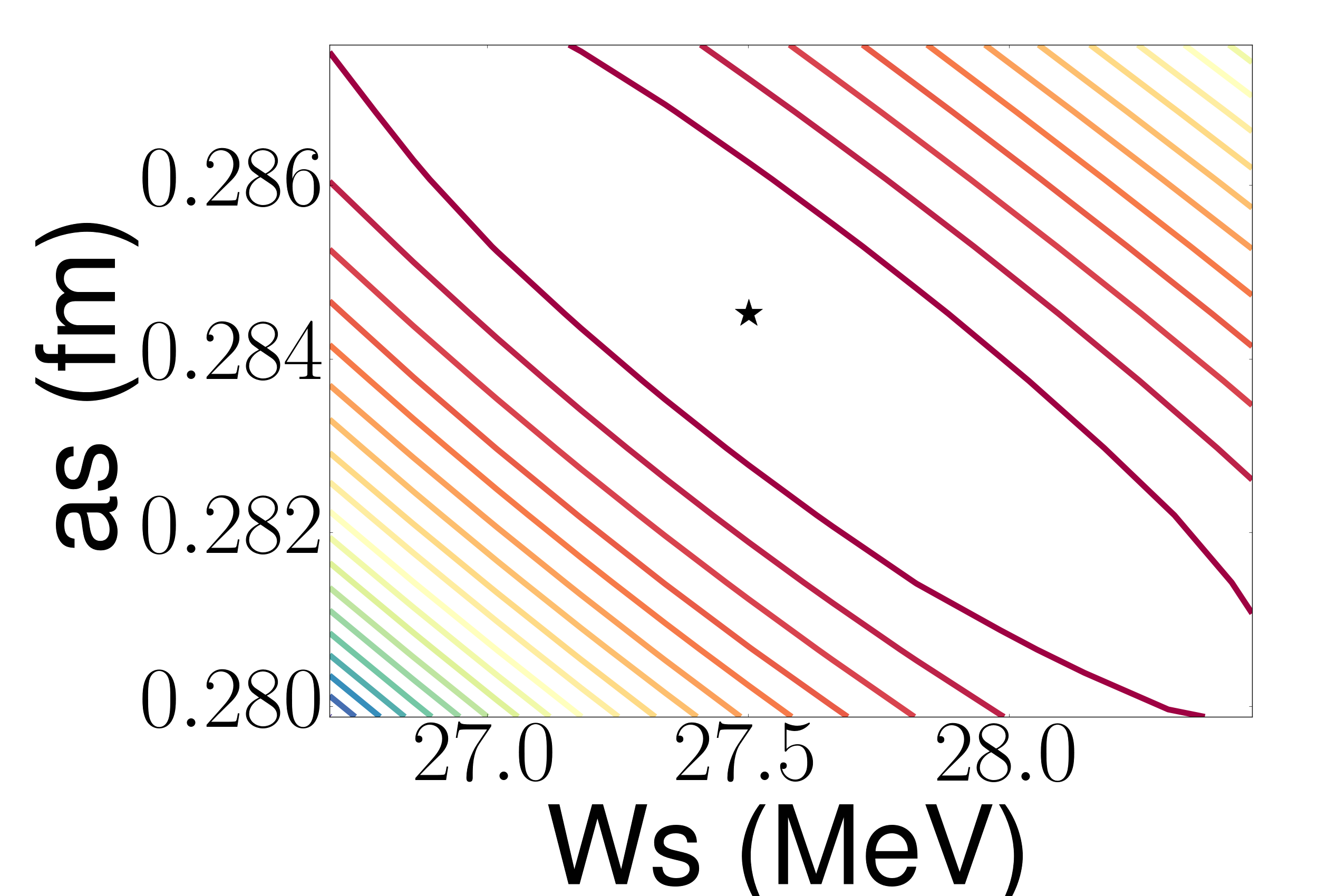} & \includegraphics[width=0.23\textwidth]{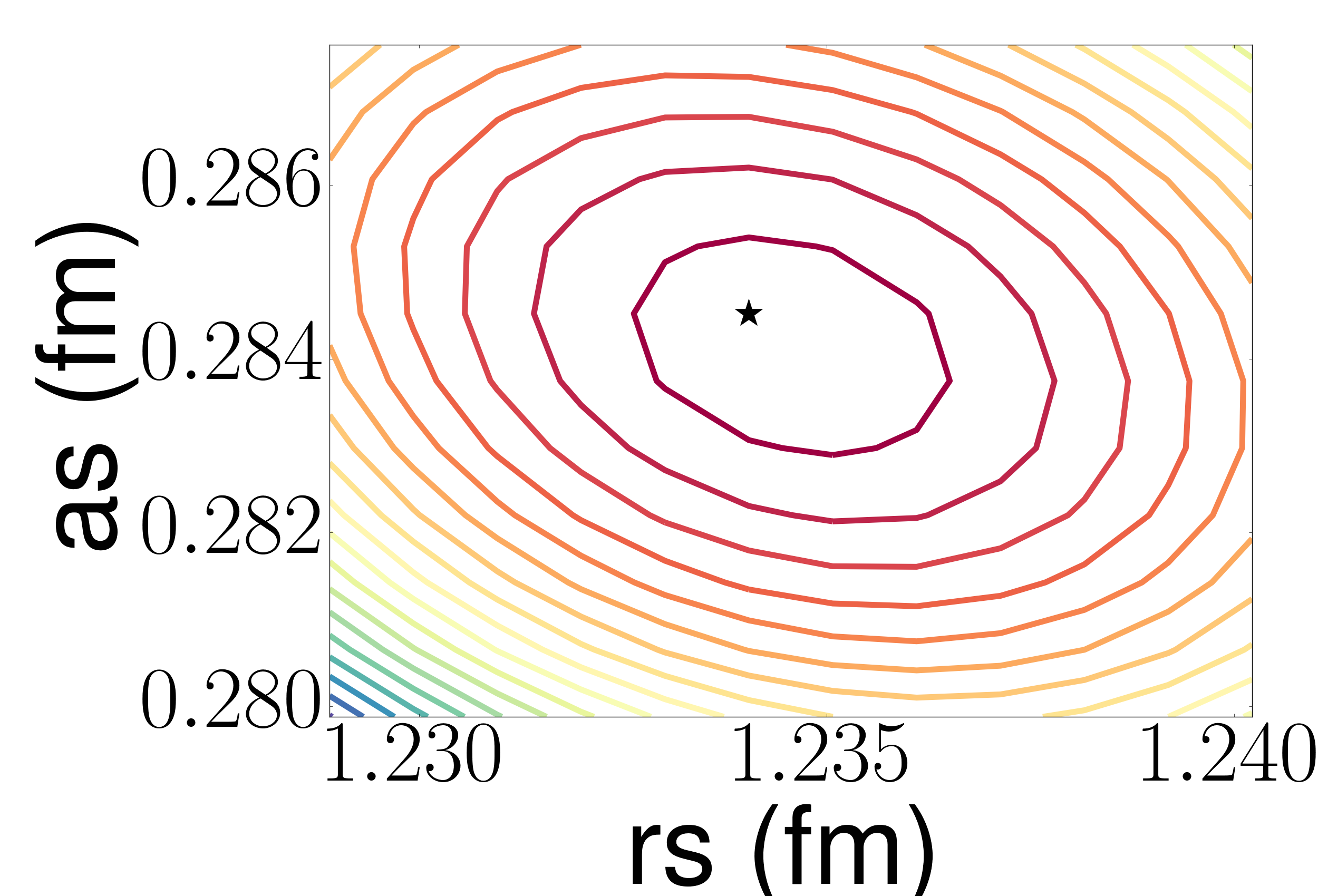} \\
\end{tabular}
\end{center}
\caption{(Color online)  Pairwise two-dimensional $\chi^2_{UC}$ contour plots for the best-fit parameterization of Table \ref{tab:uncorrfit}.  Black stars show the best-fit parameters.}
\label{fig:2Duncorr}
\end{figure}

\begin{figure}
\begin{center}
\includegraphics[width=0.5\textwidth]{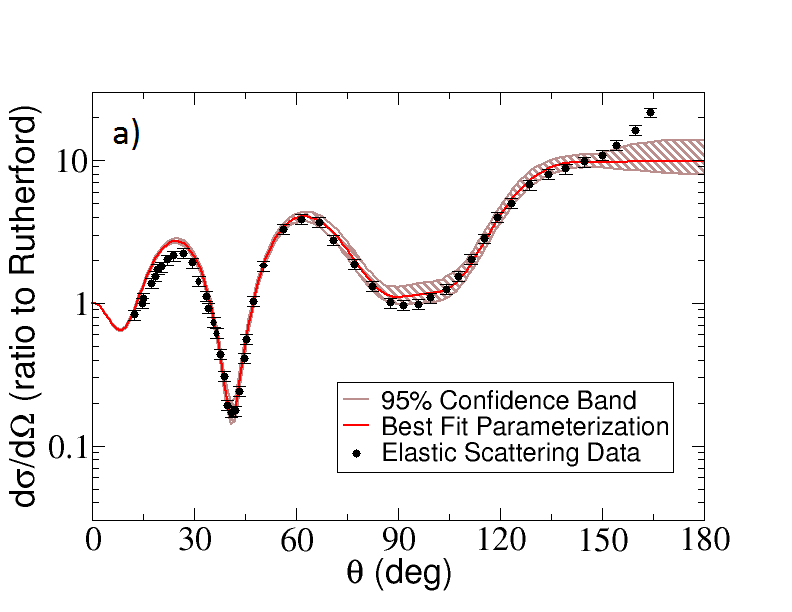}
\includegraphics[width=0.5\textwidth]{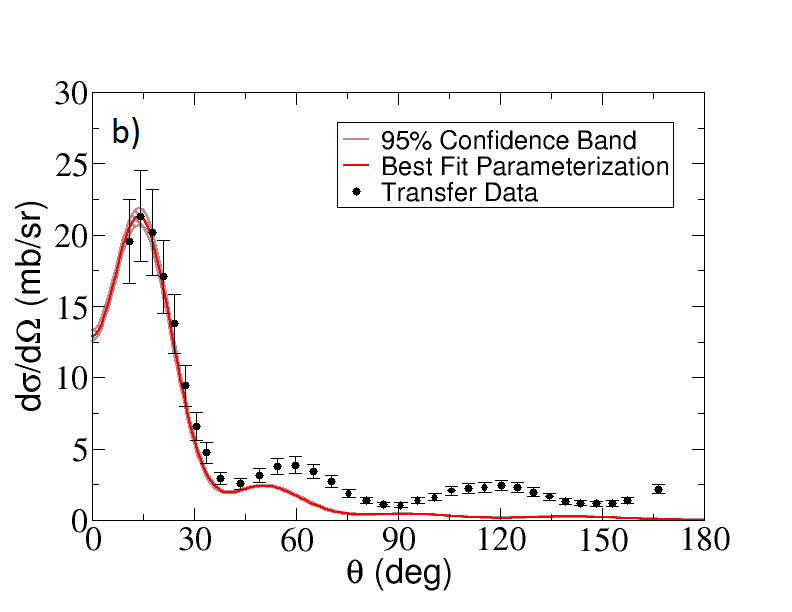}
\end{center}
\caption{(Color online)  95\% confidence bands constructed from the uncorrelated fitting of $d+^{12}$C elastic-scattering data, for elastic-scattering angular distribution (a) and the $^{12}$C(d,p)$^{13}$C(g.s.) transfer angular distribution (b, predicted) both at 11.8 MeV deuteron energy.  The cross-section calculations from the best-fit parameterization are shown in red (solid), and the 95\% confidence bands in brown (hatched); the black circles show the elastic-scattering and transfer data.}
\label{fig:CBuncorr}
\end{figure}

In order to explore the shape of the $\chi^2_{UC}$ function around this minimum, 2D $\chi^2_{UC}$ contours were constructed, shown in Figure \ref{fig:2Duncorr}, for one standard deviation on either side of the best-fit parameterization.  Since the contours are elliptical and centered around the best-fit parameterization, we can safely pull parameter sets from the multivariate Gaussian (Eq. \ref{eqn:MVG}) to construct the confidence bands.  Pulling from the multivariate Gaussian is useful because pulling directly from a $\chi^2$ distribution can be time consuming computationally.  The resulting 95\% confidence bands (brown) are shown in Figure \ref{fig:CBuncorr}.  The elastic-scattering cross section, as a ratio to the Rutherford cross section, is shown in panel (a), and the transfer cross section (normalized to the data) is shown in panel (b).  

The best-fit parameterization, shown in red, passes through most of the elastic-scattering data (Figure \ref{fig:CBuncorr}a); however, at the forward angles where the reaction models should be the most accurate, the best fit does not entirely describe the experimental cross sections.  The predicted transfer cross section (Figure \ref{fig:CBuncorr}b) has been normalized to the data, thus giving an experimental spectroscopic factor, $S^{\mathrm{exp}} = 0.435 ^{+0.019}_{-0.017}$, but the predicted angular distribution does not agree with the data for $\theta > 30 ^{\circ}$.  The average width of the elastic-scattering band is 1.2211 (as a ratio to the Rutherford cross section), and the average width of the predicted transfer cross section band is 0.75735 mb/sr.  These values seem to be extremely small and do not appear to capture the true uncertainties in the parameters.

As mentioned in Section \ref{corr}, there can be correlations in our model that are not taken into account in the fitting process.  We can visualize these model correlations by looking at the scatter of cross-section values at two different angles.  If these two angles are uncorrelated, the scatter plot should be roughly circular; a more elliptical scatter plot indicates more correlation between those two angles.  

\begin{figure}
\begin{center}
\includegraphics[width=0.5\textwidth]{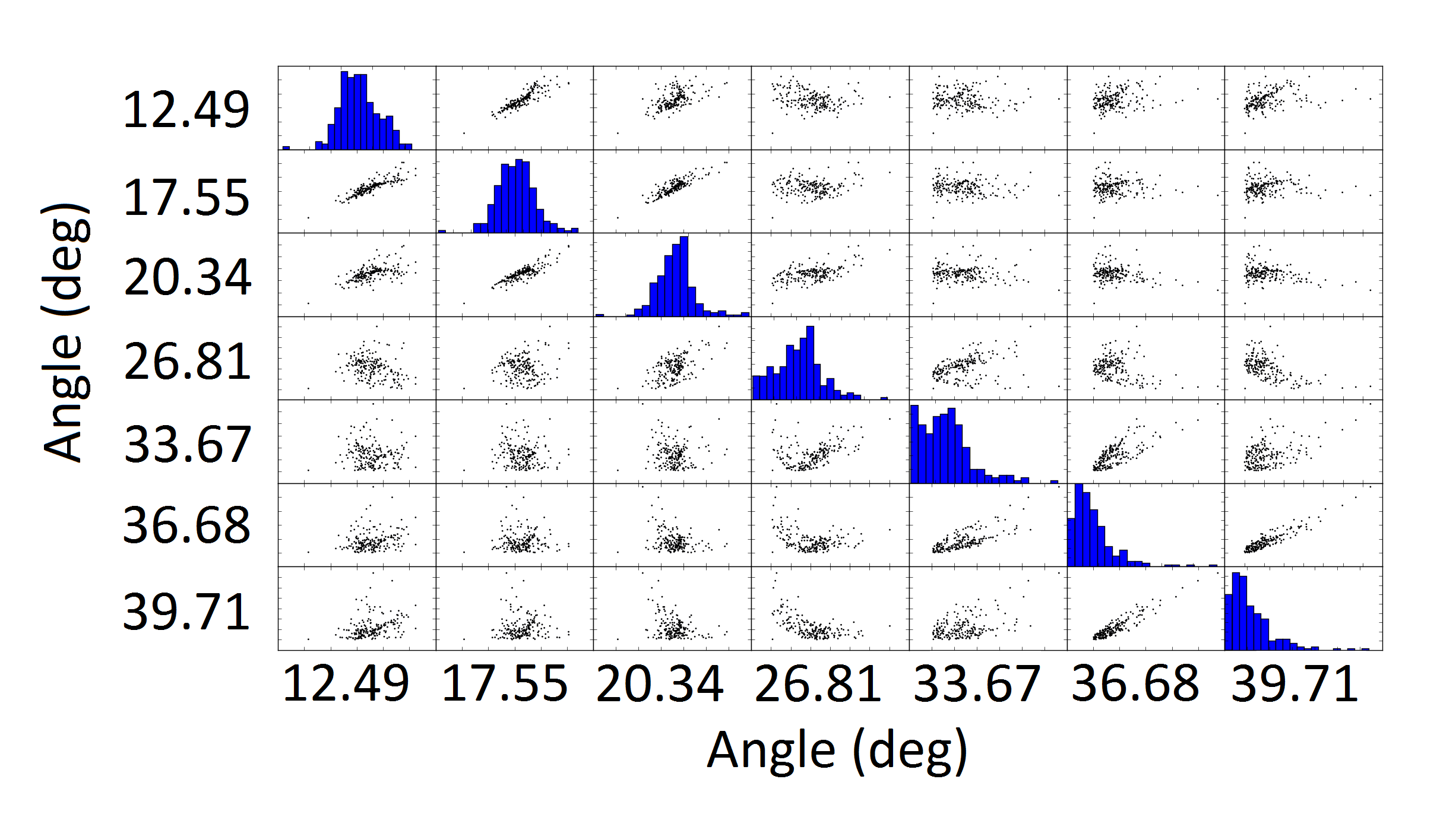}
\end{center}
\caption{(Color online)  Visual representation of model correlation for the elastic-scattering data set for $^{12}$C(d,d)$^{12}$C, for select angles.  The blue histograms show the frequency around the average value of the cross-section values at the given angle.  The scatter plots show the correlation between each pair of angles.}
\label{fig:elmodelcorr}
\end{figure}

In Figure \ref{fig:elmodelcorr}, this angular correlation is shown for a selected set of angles for the $d+^{12}$C elastic scattering we are considering here.  The histograms on the diagonal give the frequency of cross-section values at the given angle; the off-diagonal scatter plots show the angular correlations.  Angles near each other are highly correlated -- almost straight lines -- regardless of whether the angles are forward or backward, while there is a much less localized scatter for angles farther away from one another.

Because of the strong correlations, we reanalyze this case with the $\chi^2_C$ function of Eq. \ref{eqn:corrchi}.  The elastic-scattering best-fit parameterization is shown in Table \ref{tab:uncorrfit} (C).  From the differences in the parameterization, we can see that introducing correlations in the model plays a significant role in the fitting procedure, especially for the depths of the potentials.  This is discussed more in Section \ref{discussion}.

\begin{figure}
\begin{center}
\begin{tabular}{ccc}
\includegraphics[width=0.23\textwidth]{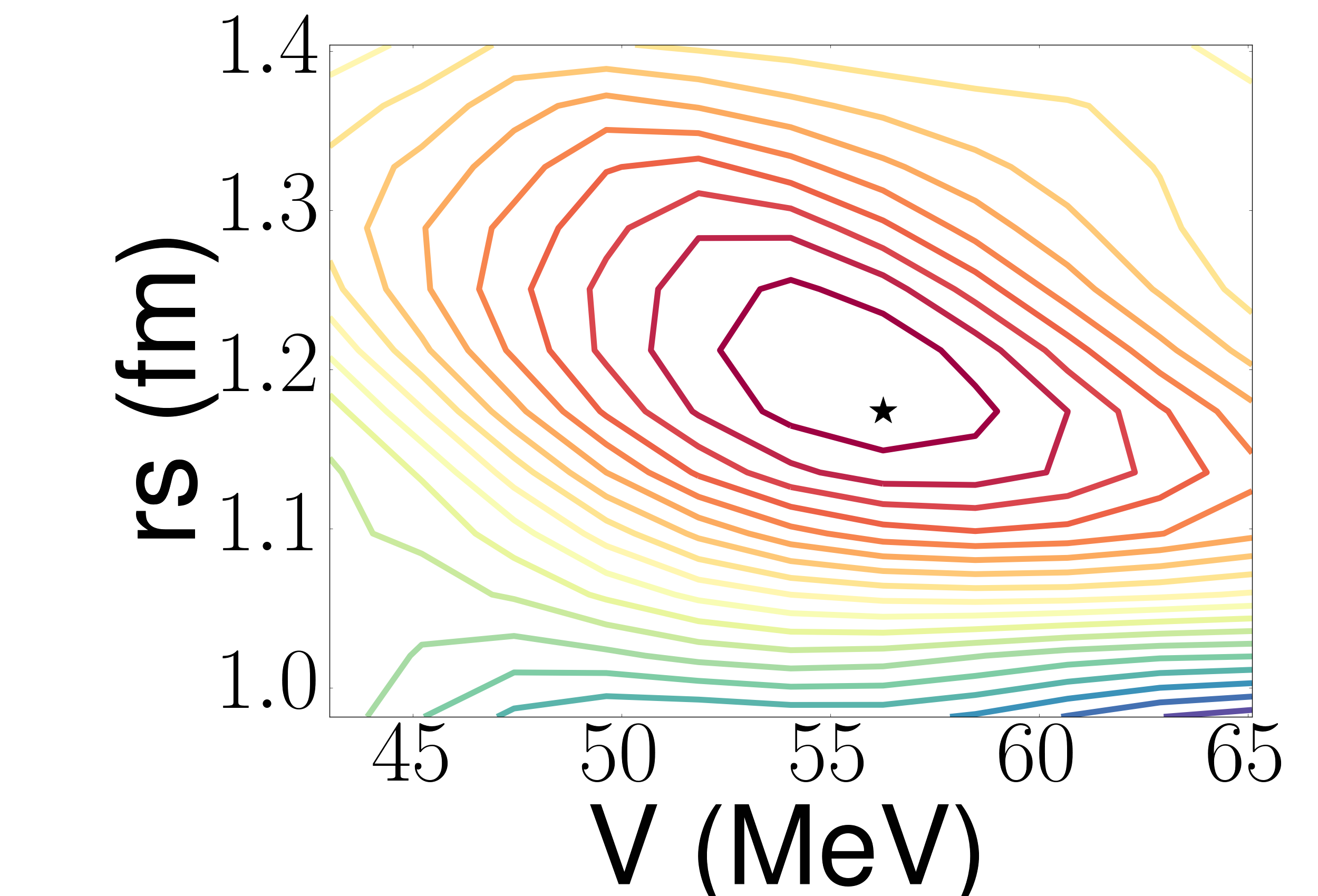} & \includegraphics[width=0.23\textwidth]{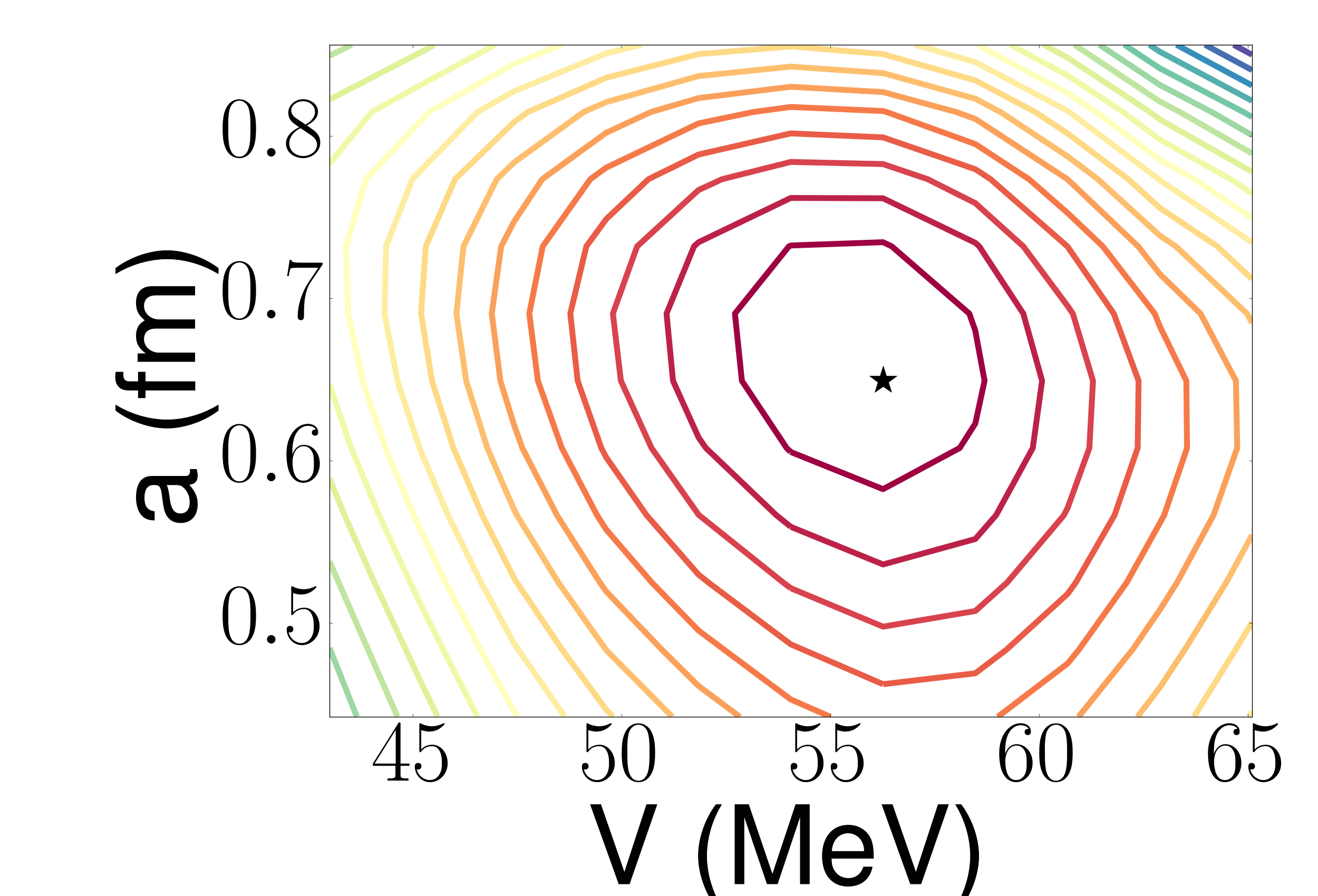} \\ 
\includegraphics[width=0.23\textwidth]{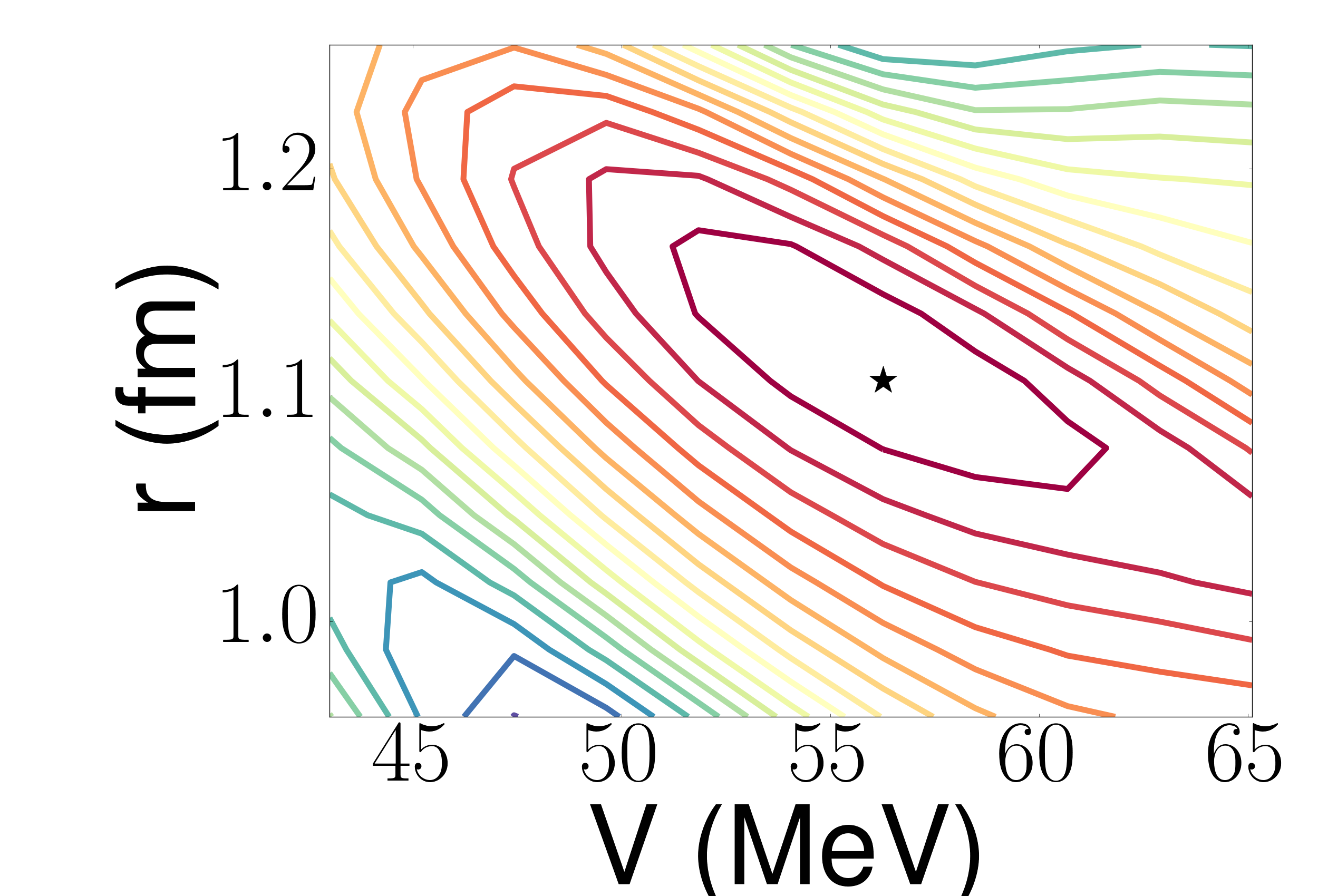} & \includegraphics[width=0.23\textwidth]{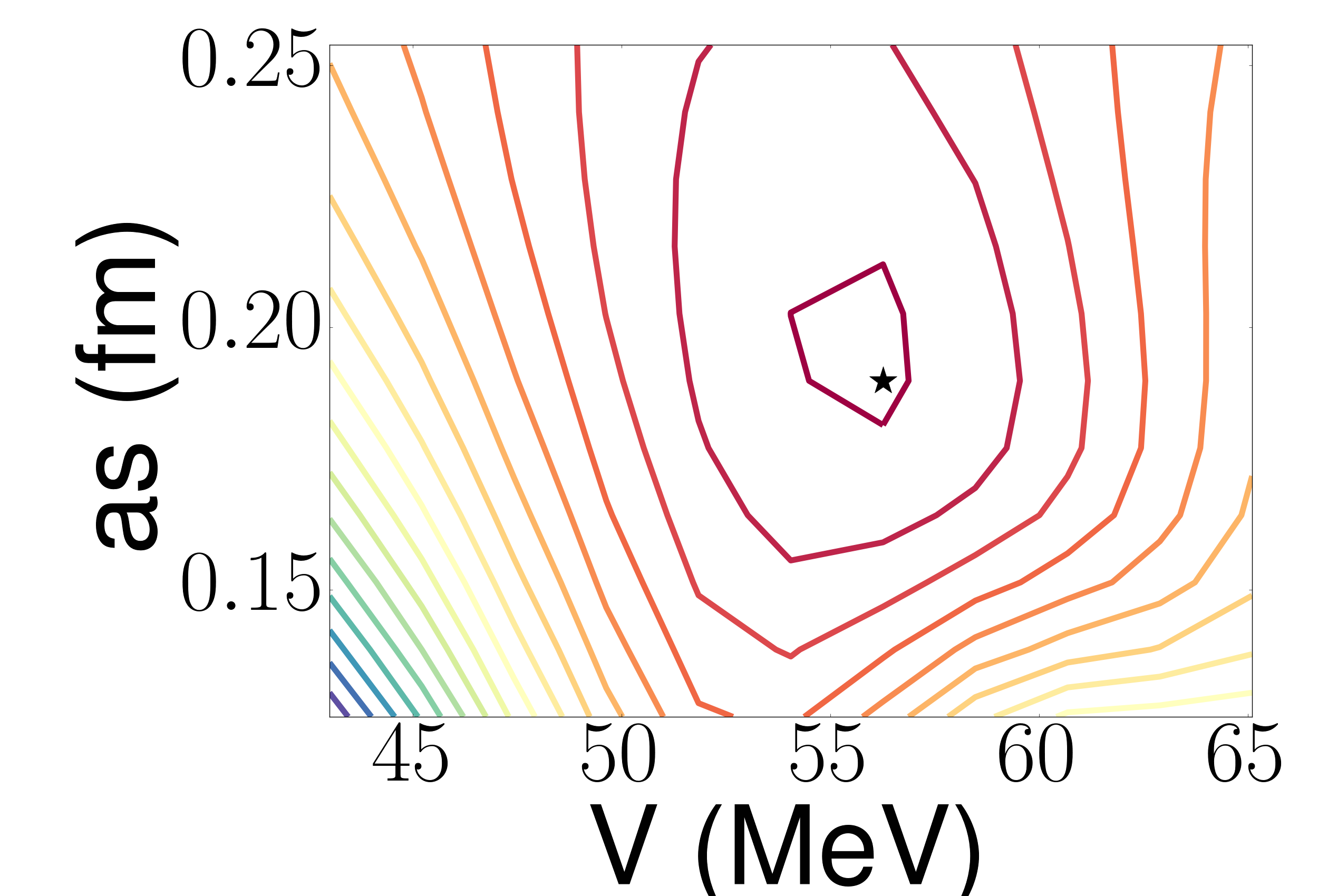} \\
\includegraphics[width=0.23\textwidth]{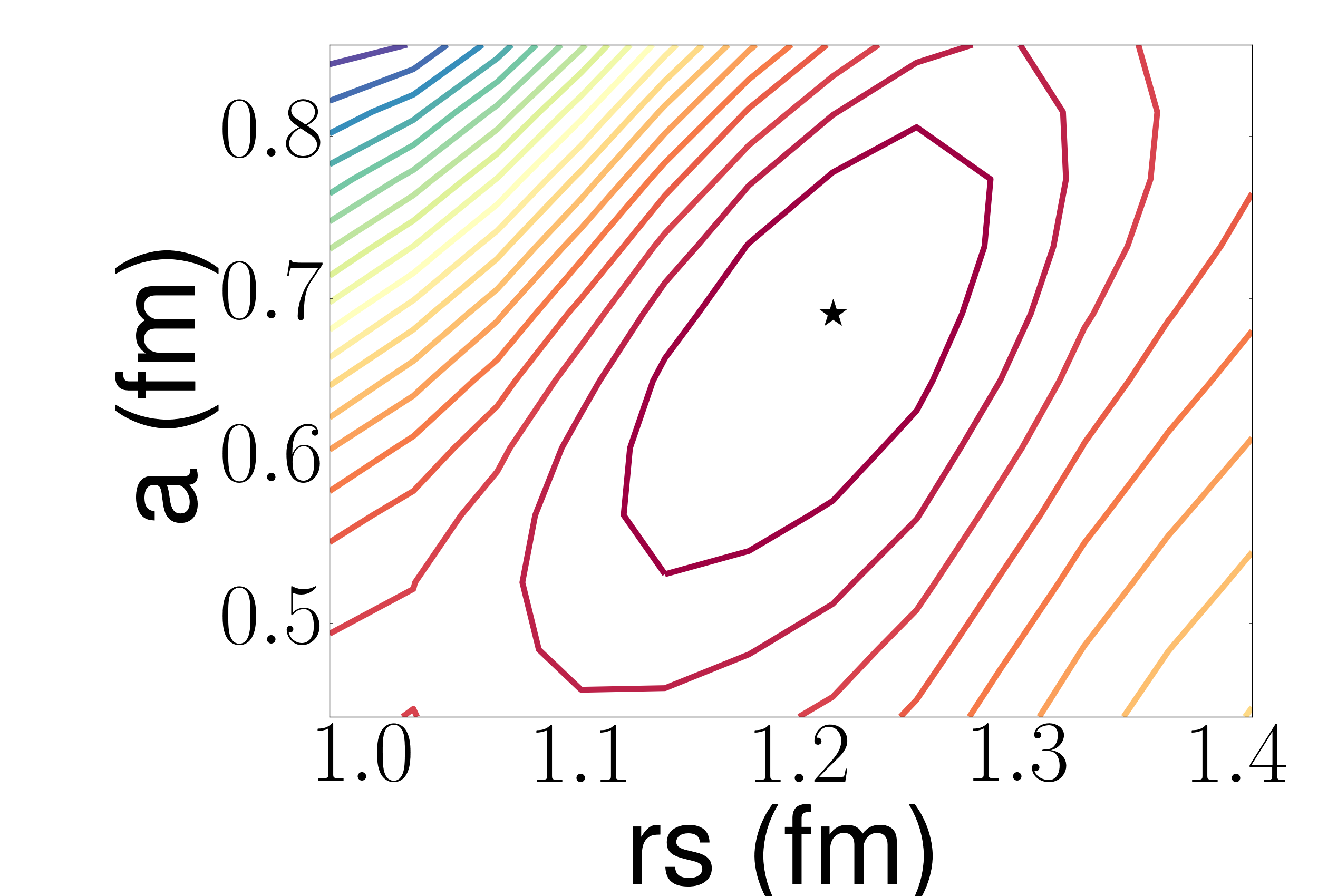} & \includegraphics[width=0.23\textwidth]{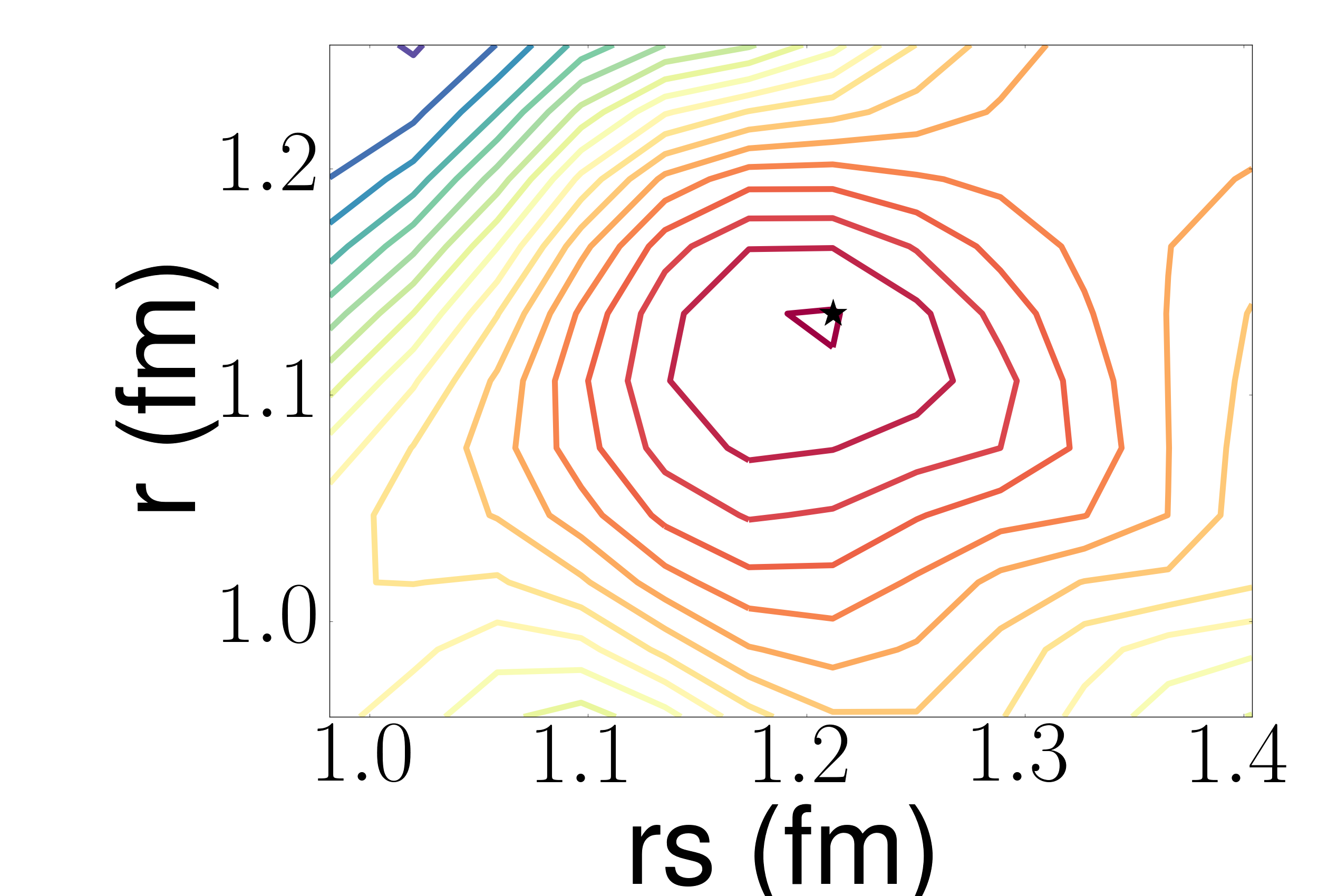} \\
\includegraphics[width=0.23\textwidth]{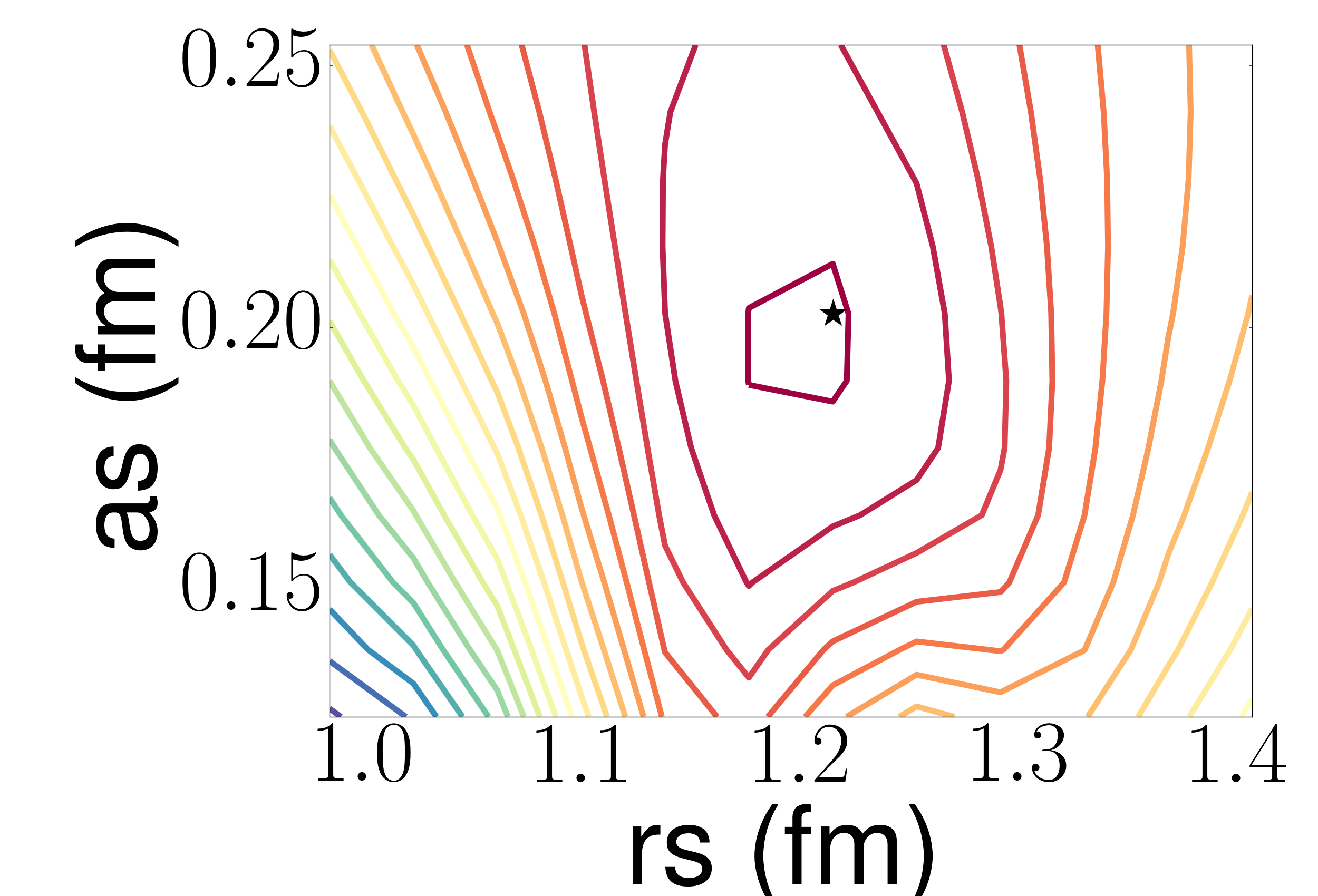} & \includegraphics[width=0.23\textwidth]{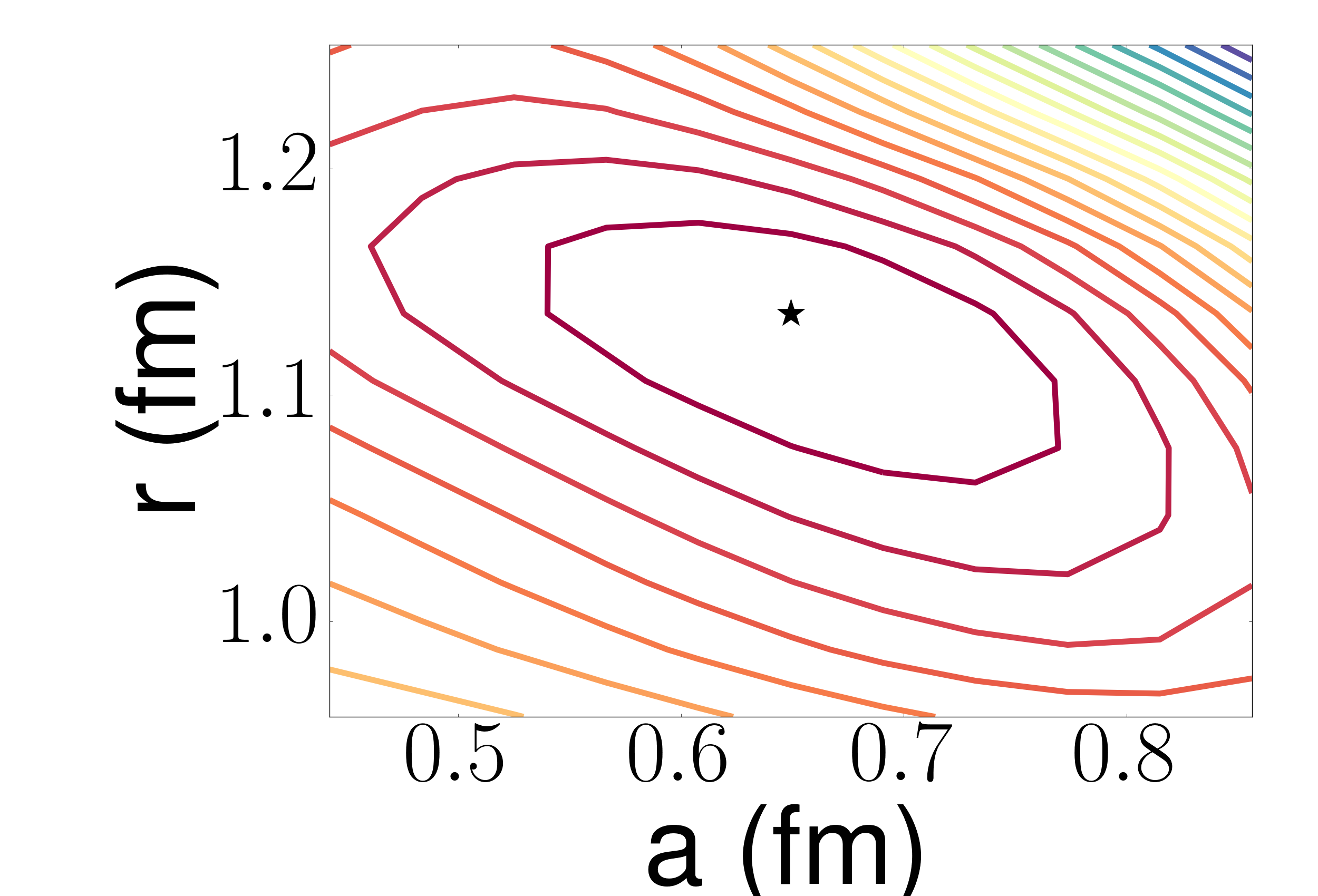} \\
\includegraphics[width=0.23\textwidth]{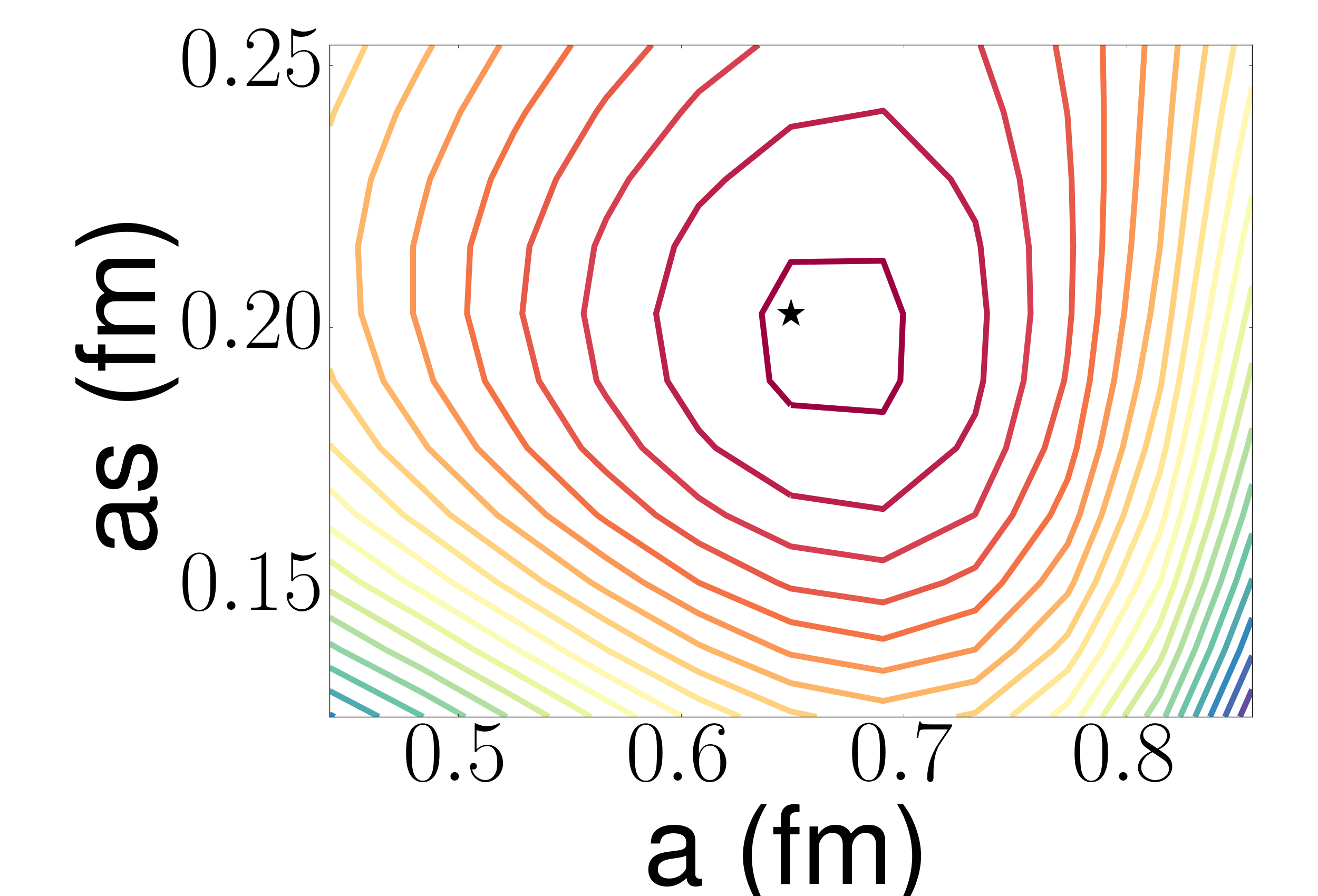} & \includegraphics[width=0.23\textwidth]{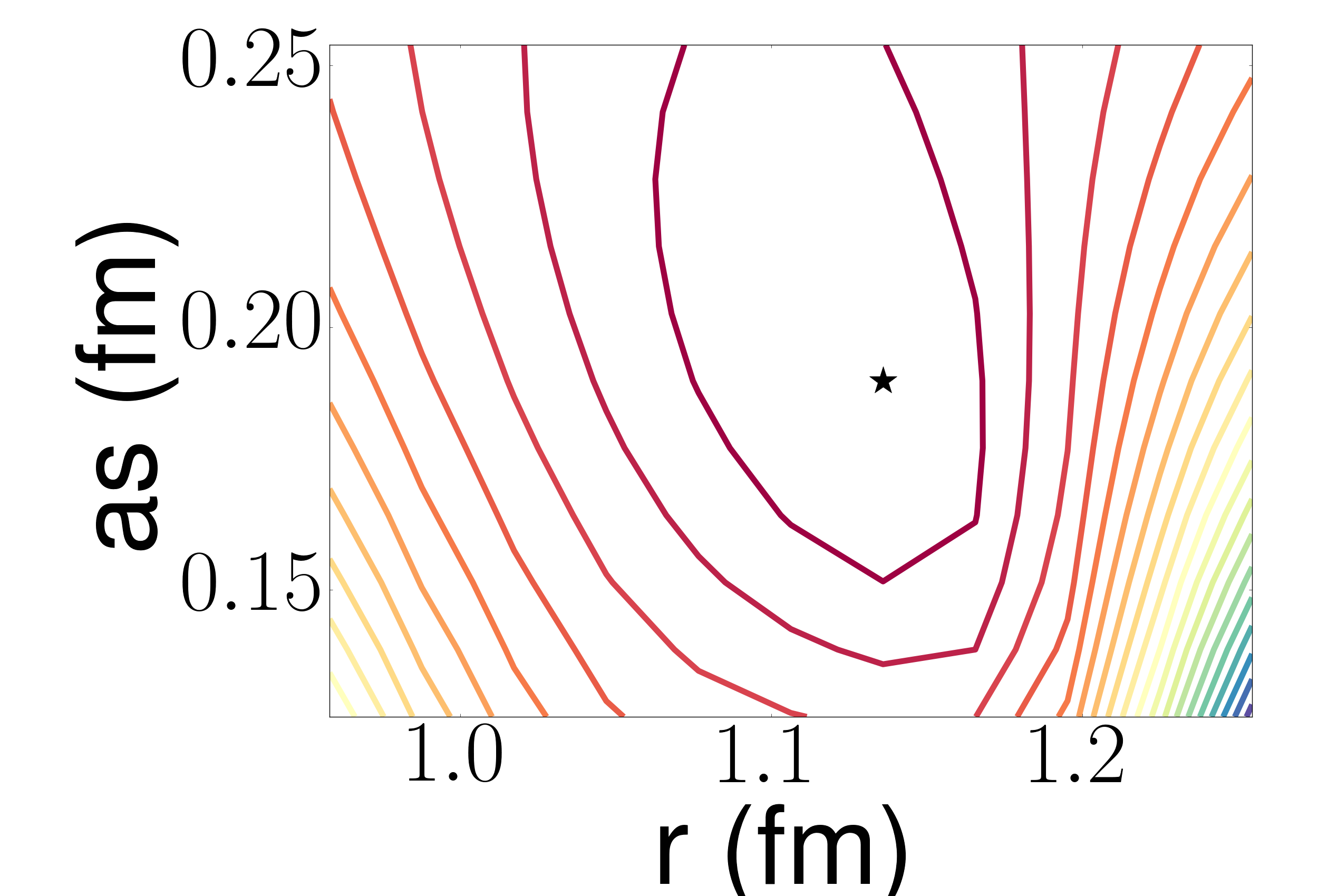} \\
\end{tabular}
\end{center}
\caption{(Color online)  Same as Figure \ref{fig:2Duncorr} for the correlated best-fit parameterization of Table \ref{tab:uncorrfit}.}
\label{fig:2Dcorr}
\end{figure}

\begin{figure}
\begin{center}
\includegraphics[width=0.5\textwidth]{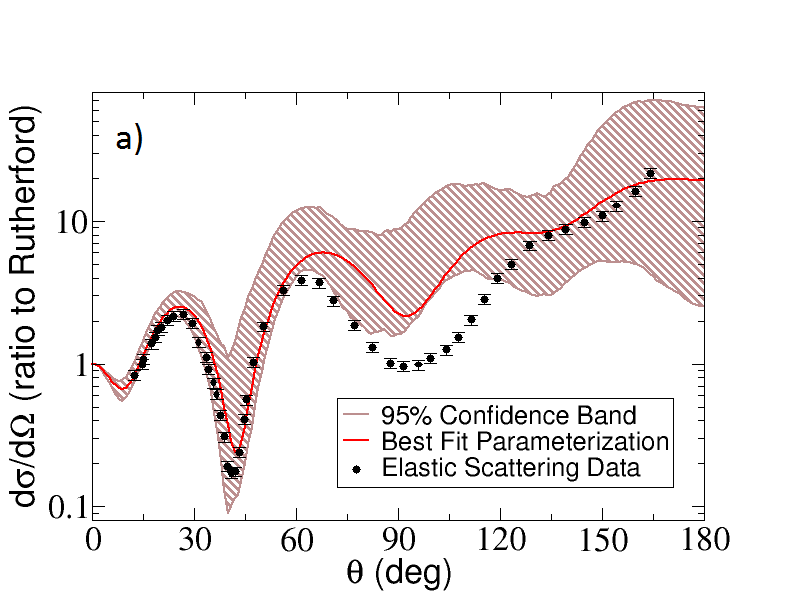}
\includegraphics[width=0.5\textwidth]{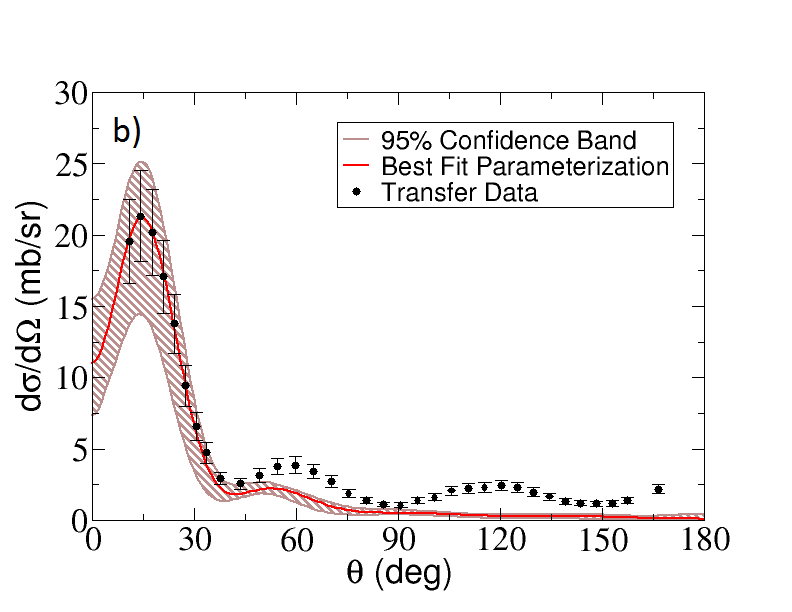}
\end{center}
\caption{(Color online)  Same as Figure \ref{fig:CBuncorr} for the correlated fit.}
\label{fig:CBcorr}
\end{figure}

The corresponding 2D $\chi^2_C$ contour plots for one standard deviation around the best-fit parameterization are shown in Figure \ref{fig:2Dcorr}.  Because the contour plots are not entirely elliptical within this region, parameters are drawn from the empirical $\chi^2_C$ distribution instead of the multivariate Gaussian.  The 95\% confidence bands for the fitted elastic-scattering angular distributions and predicted transfer angular distributions are shown in Figure \ref{fig:CBcorr}a and \ref{fig:CBcorr}b, respectively.  The elastic scattering describes the data nearly perfectly at forward angles, and even though the best-fit parameterization is above the data at intermediate angles, there is still good agreement with the overall trend as well as the magnitude at backward angles.  For the transfer calculation, the extracted spectroscopic factor, $S^{\mathrm{exp}}=0.352^{+0.223}_{-0.050}$, is smaller than for the uncorrelated case, but the predicted angular distribution improves the description of the data.  The average width for the elastic-scattering band is 16.606 (ratio to Rutherford cross section), and the average width for the predicted transfer cross section band is 6.2968 mb/sr.  These values better reflect the uncertainties of the problem.

%\begin{figure}
%\begin{center}
%\begin{tabular}{ccc}
%\includegraphics[width=0.16\textwidth]{Figures/CorrelatedExample/12CdVrs.png} & \includegraphics[width=0.16\textwidth]{Figures/CorrelatedExample/12CdVa.png} & \includegraphics[width=0.16\textwidth]{Figures/CorrelatedExample/12CdVr.png} \\
%\includegraphics[width=0.16\textwidth]{Figures/CorrelatedExample/12CdVas.png} & \includegraphics[width=0.16\textwidth]{Figures/CorrelatedExample/12Cdrsa.png} & \includegraphics[width=0.16\textwidth]{Figures/CorrelatedExample/12Cdrsr.png} \\
%\includegraphics[width=0.16\textwidth]{Figures/CorrelatedExample/12Cdrsas.png} & \includegraphics[width=0.16\textwidth]{Figures/CorrelatedExample/12Cdar.png} & \includegraphics[width=0.16\textwidth]{Figures/CorrelatedExample/12Cdaas.png} \\
% & \includegraphics[width=0.16\textwidth]{Figures/CorrelatedExample/12Cdras.png} &  \\
%\end{tabular}
%\end{center}
%\caption{(Color online)  Same as Figure \ref{fig:2Duncorr} for the best fit parameterization of Table \ref{tab:corrfit}.}
%\label{fig:2Dcorr}
%\end{figure}

\subsection{Summary of results from all calculations}
\label{summary}

We now repeat this procedure for all of the reactions listed in Table \ref{tab:cases}.  In this section, we summarize the results of these calculations, including $\chi^2$ values for each of the fits and average widths of the confidence bands.  Table \ref{tab:sumuncorr} gives this summary for the uncorrelated and correlated fits.  

\begin{table}[h]
\begin{center}
\begin{tabular}{| c | c | c | c | c |}
\hline \textbf{System}  & \textbf{$\chi^2_{UC}/M$} & \textbf{$\overline{W}_{UC}$} & \textbf{$\chi^2_C/M$} & \textbf{$\overline{W}_{C}$} \\ \hline
$^{12}$C(d,d)$^{12}$C & 4.513 & 1.3577 & 0.283 & 18.181\\
$^{12}$C(d,p)$^{13}$C(g.s.) & --- & 0.77888 & --- & 7.9226\\ \hline
$^{90}$Zr(d,d)$^{90}$Zr & 1.421 & 0.086926 & 0.142 & 0.22664 \\ 
$^{90}$Zr(d,p)$^{91}$Zr(g.s) & --- & 1.5235 & --- & 2.2567\\ \hline
$^{12}$C(n,n)$^{12}$C & 68.321 & 204.35 & 0.483 & 382.52\\
$^{12}$C(n,n')$^{12}$C(2$^+_1$) & --- & 17.212 & --- & 51.253\\ \hline
$^{48}$Ca(n,n)$^{48}$Ca & 22.344 & 134.21 & 2.142 & 380.91\\ 
$^{48}$Ca(n,n')$^{48}$Ca(2$^+_1$) & --- & 7.164 & --- & 35.586 \\ \hline
$^{54}$Fe(n,n)$^{54}$Fe & 158.098 & 151.20 & 1.080 & 92.191 \\ 
$^{54}$Fe(n,n')$^{54}$Fe(2$^+_1$) & --- & 1.4722 & --- & 2.2338 \\ \hline
$^{208}$Pb(n,n)$^{208}$Pb & 3.678 & 86.105 & 1.731 & 697.13 \\
$^{208}$Pb(n,n')$^{208}$Pb(3$^-_1$) & --- & 0.42104 & --- & 0.87376 \\ \hline
\end{tabular}
\end{center}
\caption{Summary of the properties of all of the reactions studied for this work.  The first column gives the reaction, while $\chi^2/M$ values for the uncorrelated (correlated) fits are given in column two (four), and the average width (over all angles) of the uncorrelated (correlated) 95\% confidence bands is given in the third (fifth) column.}
\label{tab:sumuncorr}
\end{table}

%\begin{table}[h]
%\begin{center}
%\begin{tabular}{| c | c | c | c |}
%\hline \textbf{System} & \textbf{E (MeV)} & \textbf{$\chi^2/m$} & \textbf{Avg. Width} \\ \hline
%$^{12}$C(d,d)$^{12}$C & 11.8 & 0.283 & 18.181 \\ 
%$^{12}$C(d,p)$^{13}$C(g.s) &  & --- & 7.9226 \\ \hline
%$^{90}$Zr(d,d)$^{90}$Zr & 12.0 & 0.142 & 0.22664\\ 
%$^{90}$Zr(d,p)$^{91}$Zr(g.s) &  & --- & 2.2567 \\ \hline
%$^{12}$C(n,n)$^{12}$C & 17.29 & 0.483 & 382.52 \\ 
%$^{12}$C(n,n')$^{12}$C(2$^+_1$) &  & --- & 51.253 \\ \hline
%$^{48}$Ca(n,n)$^{48}$Ca & 7.97 & 2.142 & 380.91 \\ 
%$^{48}$Ca(n,n')$^{48}$Ca(2$^+_1$) &  & --- & 35.586 \\ \hline
%$^{54}$Fe(n,n)$^{54}$Fe & 16.93 & 1.080 & 92.191 \\ 
%$^{54}$Fe(n,n')$^{54}$Fe(2$^+_1$) &  & --- & 2.2338 \\ \hline
%$^{208}$Pb(n,n)$^{208}$Pb & 26.0 & 1.731 & 697.13 \\ 
%$^{208}$Pb(n,n')$^{208}$Pb(3$^-_1$) &  & --- & 0.87376 \\ \hline
%\end{tabular}
%\end{center}
%\caption{Same as Table \ref{tab:sumuncorr} for the correlated fits.}
%\label{tab:sumcorr}
%\end{table}

%%%%%%%%%%%%%%%%%%%%%%%%%%%%%%%%%%%%%%%%%%%%%%%%%%%%%%%%%%%%%%%%%%%%%%%%%%%%%%%%%%%%%%%%%%%%%%%%%%%%%%%%%%%%%%%%%

\section{Discussion}
\label{discussion}

%\begin{itemize}
%\item Uncorrelated in comparison to correlated (argument for why we should be fitting with correlated fits)
%\item Parameter correlation matrices - how do they change when you include correlations (but put actual matrices in the results section?)
%\item Starting from various fits and whether or not conclusions still hold
%\item Need to say something about the percent error overall - average percent error?
%\end{itemize}

In the following section, we discuss the results of our calculations, first for the $d+^{12}$C reaction from the previous section.  We then make a few comments on the $n+^{54}$Fe and $d+^{90}$Zr cases and summarize with general comments from all of the cases studied.

\subsection{Comments on $^{12}$C(d,d)$^{12}$C and $^{12}$C(d,p)$^{13}$C}

\begin{figure}
\begin{center}
\includegraphics[width=0.5\textwidth]{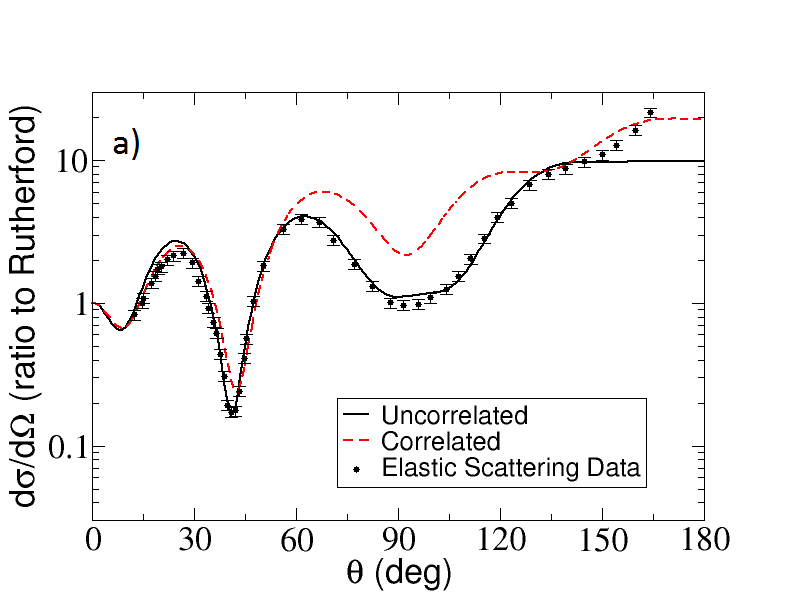}
\includegraphics[width=0.5\textwidth]{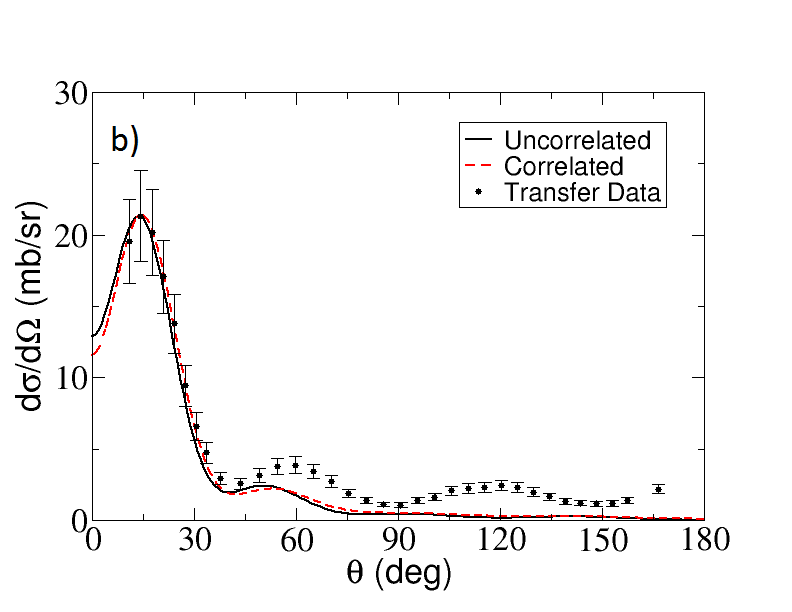}
\end{center}
\caption{(Color online)  Comparison of uncorrelated (black solid) and correlated (red dashed) cross-section calculations using the best-fit parameterization when elastic cross-section data was fit to predict transfer cross sections for $^{12}$C(d,d)$^{12}$C.  Panel (a) shows the elastic-scattering calculations, and panel (b) shows the predicted transfer calculations.}
\label{fig:elinelcomp}
\end{figure}

By looking at the comparison between the $^{12}$C(d,d)$^{12}$C best-fit elastic-scattering cross sections and the $^{12}$C(d,p)$^{13}$C(g.s.) transfer predictions, we can better understand the differences that arise when model correlations are included in the calculation.  Figure \ref{fig:elinelcomp} shows this comparison between the best-fit parameterizations of the uncorrelated (black solid) and correlated (red dashed) angular distributions for elastic-scattering (panel a) and transfer (panel b) cross sections.  With the model correlations, the angular distributions from the best-fit parameterization provide an overall more consistent description of the angular distributions.  For the elastic scattering, this is true more so around grazing angles, which is where we expect our model to be the most accurate.  Even at backward angles, however, the experimental elastic-scattering angular distribution is better described by the correlated angular distribution, which continues to rise instead of flattening off, as in the uncorrelated calculation.  At central angles (around $\sim 90 ^{\circ}$), the uncorrelated calculation is in almost perfect agreement with the data, but the correlated calculation still reproduces the overall trend.

Furthermore, $\chi^2_{C}$ is lower than $\chi^2_{UC}$ by a factor of about 8.  Correlated fitting also produces larger confidence bands for both the elastic and transfer cross-section calculations, as can be seen by comparing the third and fifth columns of Table \ref{tab:sumuncorr}.

Table \ref{tab:uncorrfit} shows significant differences between the correlated and uncorrelated parameterizations.  The real and imaginary potential depths for the correlated minimum are extremely atypical, which could explain the small spectroscopic factor that is extracted from the calculation.  The uncorrelated spectroscopic factor still falls within the error bands of the correlated calculation; within error bars, these two parameterizations are consistent.  If we instead fix the real volume depth at the more physical value of 111.505 MeV and vary the remaining five parameters to find a minimum, the imaginary surface depth increases to keep a low $\chi^2_C$ value, and the extracted spectroscopic factor does not change.  

There is an understanding that DWBA does not provide an accurate description of this reaction (e.g., \cite{Delaunay2005}) and that an effective deuteron optical potential that does not explicitly account for $np$ breakup is unreliable \cite{Nunes2011}.  The strong variation of the minimum found between the uncorrelated and correlated cases can be a symptom of the reaction model simplification.

\subsection{Comments on other specific reactions}

\subsubsection{$n+^{54}$Fe elastic and inelastic scattering}

\begin{figure}
\begin{center}
\includegraphics[width=0.5\textwidth]{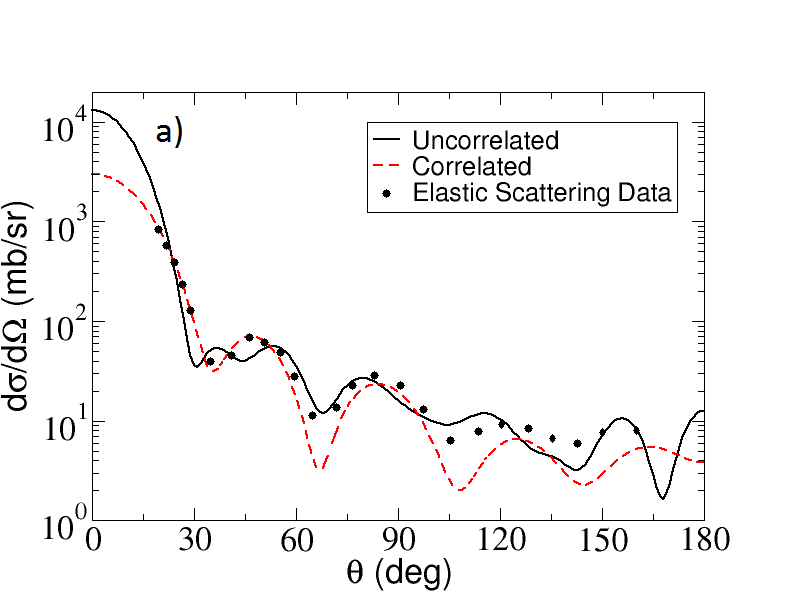}
\includegraphics[width=0.5\textwidth]{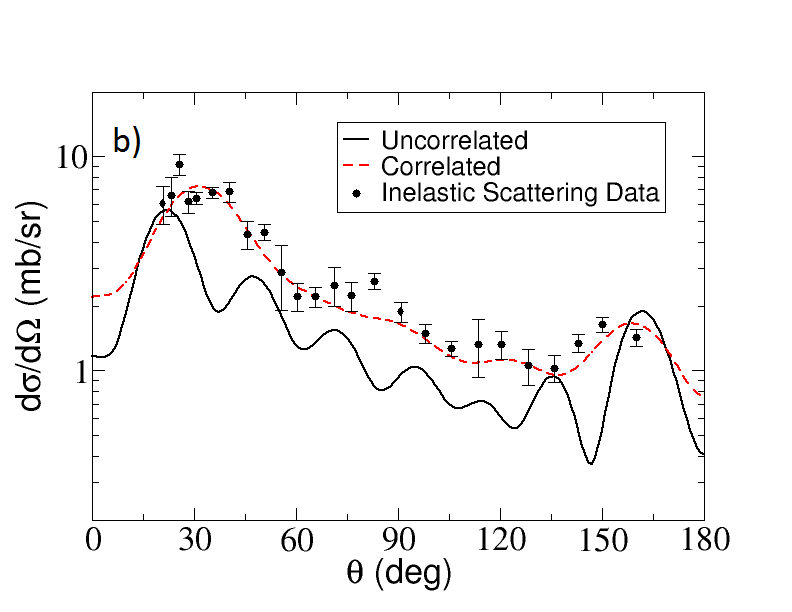}
\end{center}
\caption{(Color online)  Comparison of correlated and uncorrelated cross-section calculations using the best-fit parameterization when elastic-scattering data was fit to predict inelastic cross sections, for $^{54}$Fe-n scattering.  Panel (a) shows the elastic-scattering fits, and panel (b) shows the inelastic-scattering predictions.  Shown are uncorrelated (black solid) and correlated (red dashed) fits and predictions.}
\label{fig:54Fecomp}
\end{figure}

Some of the reactions we studied show a greater difference between the cross sections resulting from the uncorrelated and correlated minima.  Figure \ref{fig:54Fecomp} shows the angular distributions for elastic scattering (a) and inelastic scattering (b) compared with data for $n+^{54}$Fe at 16.93 MeV.  The correlated best-fit calculations (red, dashed) better describe the trends of the experimental angular distributions compared to the uncorrelated calculations (black, solid) for the elastic calculation.  Our results also show that the correlated best-fit prediction better describes both the shape and the magnitude of the experimental inelastic angular distributions.

The best-fit parameters for these uncorrelated and correlated minima for $n+^{54}$Fe scattering are given in Table \ref{tab:54Femin}.  Despite the small radius for the real volume term, the rest of the parameters for the correlated minimum take on more physically meaningful values than the corresponding parameters from the uncorrelated minimum; this is especially evident for the depths of the potentials.

\begin{table}
\begin{center}
\begin{tabular}{|c|c|c|c|c|c|c|}
\hline & \textbf{V$_\mathrm{o}$(MeV)} & \textbf{r$_\mathrm{o}$(fm)} & \textbf{a$_\mathrm{o}$(fm)} & \textbf{W$_\mathrm{V}$(MeV)} & \textbf{r$_\mathrm{w}$(fm)} & \textbf{a$_\mathrm{w}$(fm)} \\ \hline
UC & \IB{29.411} & \IG{1.609} & \IB{0.4439} & \IB{7.432} & \IG{1.078} & \IG{0.5603} \\ \hline
C & \IB{47.371} & \IG{0.9324} & \IB{0.6001} & \IB{2.292} & \IG{1.161} & \IG{0.1120} \\ \hline
 & --- & --- & --- & \textbf{W$_\mathrm{s}$(MeV)} & \textbf{r$_\mathrm{s}$(fm)} & \textbf{a$_\mathrm{s}$(fm)} \\ \hline
UC & --- & --- & --- & \IB{22.550} & \IB{1.504} & \IG{0.1246} \\ \hline
C & --- & --- & --- & \IB{5.433} & \IB{1.104} & \IG{0.5852} \\ \hline
\end{tabular}
\end{center}
\caption{(Color online)  Best-fit parameters for $^{54}$Fe(n,n)$^{54}$Fe elastic scattering.  The second and fifth (third and sixth) rows give the uncorrelated (correlated) minimum.  V$_\mathrm{o}$, r$_\mathrm{o}$, and a$_\mathrm{o}$ are the real volume terms, W$_\mathrm{V}$, r$_\mathrm{w}$, and a$_\mathrm{w}$ are the imaginary volume terms, and W$_\mathrm{s}$, r$_\mathrm{s}$, and a$_\mathrm{s}$ are the imaginary surface terms.}
\label{tab:54Femin}
\end{table}

\subsubsection{$d+^{90}$Zr elastic scattering and transfer}

The spectroscopic factors for $^{90}$Zr(d,p)$^{91}$Zr(g.s.) are $0.720^{+0.097}_{-0.060}$ for the uncorrelated parameterization and $0.689 ^{+0.194}_{-0.079}$ for the correlated parameterization, which are less than the values extracted in \cite{90Zrdp}.  For this case, the parameterizations between the uncorrelated and correlated minima are significantly different, but this does not have much of an effect on the resulting transfer angular distributions, as evident by the similarities in the spectroscopic factors.  Again, the resulting 95\% confidence band is larger for the correlated calculation.

\subsection{General comments}
\label{gencomments}

We now can discuss general properties for all of the reactions that we studied. In all cases, the correlated minima provide better descriptions of the data at forward angles, if not everywhere.  As shown in the preceding sections, this is true for both fitted and predicted cross sections.   When model correlations are present, however, the best-fit calculations may vary systematically from the data.  Moreover, $\chi^2_C$ is always smaller than $\chi^2_{UC}$, because of the introduction of the model covariance matrix.  

In almost all cases, the average width of the confidence bands from the correlated fit is larger than for the uncorrelated case.  Part of the reason for this is that the parameter covariance matrix is larger for the correlated fits than for the uncorrelated fits (because of the introduction of the model covariance matrix) -- and even though it is scaled by $s^2$ as defined in Eq. \ref{eqn:ssq}, there is still a wider range for the parameters to be pulled from.  

Although not shown, we can make two comments on the correlations between the fitted parameters across all reactions studied here.  The first is that radii tend to decouple by 10--15\% from the potential depth when going from uncorrelated to correlated fitting.  (These are generally the most correlated pairs of parameters.)  Second, the depths of the interactions couple more strongly to one another by 30--40\% when model correlations are added.  This coupling may influence the wider confidence bands for the correlated case, depending on whether the couplings are positively or negatively correlated.

These same conclusions also hold when starting from other global optical model parameterizations and performing an independent minimization.  Even with different spin-orbit potential parameters and different Coulomb radii, the general comments of this section are valid.  

We conclude that a large uncertainty can be introduced from the variation in parameters.  However, this comes from assuming nearly nothing about constraints on the parameter space; including more information about the model space before the fit could decrease this uncertainty.  Furthermore, there are still large modeling errors that are not treated here; these also need to be taken into account systematically.

%%%%%%%%%%%%%%%%%%%%%%%%%%%%%%%%%%%%%%%%%%%%%%%%%%%%%%%%%%%%%%%%%%%%%%%%%%%%%%%%%%%%%%%%%%%%%%%%%%%%%%%%%%%%%%%%%

\section{Conclusions and future work}
\label{conclusions}

%\begin{itemize}
%\item Summary of what we've found
%\item Ways to proceed in the future (Bayesian - better constrain confidence bands?, evaluating the quality of a model/potential without fitting, extrapolating to systems without data)
%\end{itemize}

We have used a statistical model to construct 95\% confidence bands for six reactions in the range $A$=12--208 at energies below 30 MeV/u in order to study uncertainties coming from the parameterization of the optical potentials.  These parameters were allowed to vary in order to fit elastic cross sections to elastic-scattering data and were then used to predict cross sections for inelastic scattering and transfer reactions.  A correlated $\chi^2$ function was introduced to take into account some of the correlations present in the reaction model.  One case, fitting $^{12}$C(d,d)$^{12}$C elastic scattering to predict $^{12}$C(d,p)$^{13}$C cross sections, was discussed in detail.

In general, we find that the $\chi^2_C$ function provides a more physical description of the cross section, in terms of parameter values in the optical potentials and the shape of the calculated angular distributions with respect to the experimental ones.  The $\chi^2_C$ values are lower than the $\chi^2_{UC}$ values.  However, the 95\% confidence bands constructed from the correlated fits are larger than the bands constructed from the uncorrelated fits.  Although the optical potential parameters are all highly coupled, in the correlated calculations, the potential depths and corresponding radii decouple slightly from one another, but the potential depths couple more strongly to one another.

The incoming elastic channel is only one part of the optical potential that must be specified.  For the transfer cases, ambiguity in the outgoing channels also leads to uncertainties in reaction observables, which could be systematically studied if elastic-scattering data for both the incoming and outgoing channels were available.  Having elastic, inelastic, and transfer data for various isotopes across an isotopic chain at several energies would allow us to study systematic trends of these confidence bands in order to better understand how our predictive power changes toward the edges of stability and towards the edges of the known nuclear chart.  

A better description of the uncertainties coming from the parameter variations is only half of the story.  Uncertainties coming from the approximations within the theory framework must also be quantified, and these effects will likely be dependent on the specific reaction being studied.  Couplings to higher lying excited states in the target nucleus of transfer reactions can change the magnitude of the cross section at the peak by up to 15\% \cite{Delaunay2005,Nunes2001}.  Inclusion of deuteron break-up is another important effect, and even when using the adiabatic wave approximation, differences between results from those calculations and full three-body Faddeev calculations can be around 20\% \cite{Nunes2011a}.  For these reasons, depending on the reaction model used, one will obtain differing spectroscopic factors extracted from the same transfer data \cite{Deltuva2016}.  However, by including more degrees of freedom into the reaction model than what we have done in the present work, error bands will potentially decrease as more reaction channels can be described and therefore more constraints can be added to the fitting procedure.  Investigations into model uncertainties are underway.  

%%%%%%%%%%%%%%%%%%%%%%%%%%%%%%%%%%%%%%%%%%%%%%%%%%%%%%%%%%%%%%%%%%%%%%%%%%%%%%%%%%%%%%%%%%%%%%%%%

\begin{center}
\textbf{ACKNOWLEDGMENTS}
\end{center}

%\begin{itemize}
%\item ANL grants
%\end{itemize}

This work was supported by the Stewardship Science Graduate Fellowship program under Grant No. DE-NA0002135.  This work was also supported by the National Science Foundation under Grant Nos. PHY-1403906 and PHY-1520929, under the auspices of the Department of Energy under Contract No. DE-FG52-08NA28552, and by the US Department of Energy, Office of Science, Advanced Scientific Computing Research, under Contract No. DE-AC02-06CH11357.

%%%%%%%%%%%%%%%%%%%%%%%%%%%%%%%%%%%%%%%%%%%%%%%%%%%%%%%%%%%%%%%%%%%%%%%%%%%%%%%%%%%%%%%%%%%%%%%%%

\appendix

%%%%%%%%%%%%%%%%%%%%%%%%%%%%%%%%%%%%%%%%%%%%%%%%%%%%%%%%%%%%%%%%%%%%%%%%%%%%%%%%%%%%%%%%%%%%%%%%%

%%%%%%%%%%%%%%%%%%%%%%%%%%%%%%%%%%%%%%%%%%%%%%%%%%%%%%%%%%%%%%%%%%%%%%%%%%%%%%%%%%%%%%%%%%%%%%%%%

\bibliography{UQbands}

\end{document}